\documentclass[11pt,3p,times]{elsarticle}

\usepackage{amsmath,amsfonts,amssymb}
\usepackage{amsthm}
\usepackage{bm}
\usepackage{graphicx,epstopdf}
\usepackage[position=top,labelfont=normalfont,textfont=normalfont,singlelinecheck=off,justification=raggedright]{subfig}
\usepackage{floatrow}
\usepackage[dvipsnames]{xcolor}
\usepackage{setspace}
\usepackage{enumerate,etaremune}
\usepackage{booktabs}
\usepackage{tabularx,multirow}
\usepackage{framed}
\usepackage{hhline}
\usepackage{url}
\usepackage[unicode,bookmarks=false]{hyperref}
\hypersetup{
  colorlinks,
  citecolor=Blue,linkcolor=Blue,urlcolor=Blue
}

\usepackage{lineno}
\usepackage[textsize=footnotesize,linecolor=red,backgroundcolor=red!25,bordercolor=red]
  {todonotes}

\usepackage{tikz}
\usetikzlibrary{shapes.geometric}
\usetikzlibrary{shapes,arrows}
\usetikzlibrary{plotmarks}

\usetikzlibrary{calc}

\usepackage{physics}

\usepackage{algorithm}
\usepackage{algorithmicx,algpseudocode}
\usepackage{cases}

\newcommand{\eg}{{\it e.g.}}
\newcommand{\ie}{{\it i.e.}}
\newcommand{\cf}{{\it cf.}}
\newcommand{\etal}{{\it et al.}}
\newcommand{\tensor}[1]{\bm{#1}}
\newcommand{\stress}{\sigma}

\newcommand{\tstress}{\tensor{\stress}}

\newcommand{\pd}{\partial}

\newcommand{\el}{\mathrm{e}}

\newcommand{\rn}[1]{\uppercase\expandafter{\romannumeral #1\relax}}
\newcommand{\cn}{\mathrm{N}}

\DeclareMathOperator{\dyadic}{\otimes}
\newsavebox{\dotbox}

\theoremstyle{remark}

\newcommand{\revised}[1]{{\color{black} #1}}

\newcolumntype{L}[1]{>{\raggedright\let\newline\\arraybackslash\hspace{0pt}}m{#1}}
\newcolumntype{C}[1]{>{\centering\let\newline\\arraybackslash\hspace{0pt}}m{#1}}
\newcolumntype{R}[1]{>{\raggedleft\let\newline\\arraybackslash\hspace{0pt}}m{#1}}

\allowdisplaybreaks

\biboptions{sort&compress,square,comma,numbers}
\AtBeginDocument{\hypersetup{citecolor=MidnightBlue,linkcolor=MidnightBlue,urlcolor=MidnightBlue}}

\usepackage{etoolbox}
\makeatletter
\patchcmd{\ps@pprintTitle}
{Preprint submitted to}
{~}
{}{}
\makeatother

\begin{document}

\begin{frontmatter}

\title{A phase-field model for quasi-dynamic nucleation, growth, and propagation of rate-and-state faults}

\author[LLNL]{Fan Fei}
\author[UIUC_ME,UIUC_CEE]{Md Shumon Mia}
\author[UIUC_CEE,UIUC_BECKMAN]{Ahmed E. Elbanna\corref{corr}}
\ead{elbanna2@illinois.edu}
\author[KAIST]{Jinhyun Choo\corref{corr}}
\ead{jinhyun.choo@kaist.ac.kr}

\cortext[corr]{Corresponding Authors}
\address[LLNL]{Atmospheric, Earth, and Energy Division, Lawrence Livermore National Laboratory, Livermore, CA, USA}
\address[UIUC_ME]{Department of Mechanical Science and Engineering, University of Illinois at Urbana-Champaign, Urbana, IL, USA}
\address[UIUC_CEE]{Department of Civil and Environmental Engineering, University of Illinois at Urbana-Champaign, Urbana, IL, USA}
\address[UIUC_BECKMAN]{Beckman Institute of Advanced Science and Technology, University of Illinois at Urbana-Champaign, Urbana, IL, USA}
\address[KAIST]{Department of Civil and Environmental Engineering, KAIST, Daejeon, South Korea}

\journal{~}


\begin{abstract}
Despite its critical role in the study of earthquake processes, numerical simulation of the entire stages of fault rupture remains a formidable task.
The main challenges in simulating a fault rupture process include complex evolution of fault geometry, frictional contact, and off-fault damage over a wide range of spatial and temporal scales.
Here, we develop a phase-field model for quasi-dynamic fault nucleation, growth, and propagation, which features two standout advantages: (i) it does not require any sophisticated algorithms to represent fault geometry and its evolution; and (ii) it allows for modeling fault nucleation, propagation, and off-fault damage processes with a single formulation. 
Built on a recently developed phase-field framework for shear fractures with frictional contact, the proposed formulation incorporates rate- and state-dependent friction, radiation damping, and their impacts on fault mechanics and off-fault damage.  
We show that the numerical results of the phase-field model are consistent with those obtained from  well-verified approaches that model the fault as a surface of discontinuity, without suffering from the mesh convergence issue in the existing continuous approaches to fault rupture (\eg~the stress glut method). 
Further, through numerical examples of fault propagation in various settings, we demonstrate that the phase-field approach may open new opportunities for investigating complex earthquake processes that have remained overly challenging for the existing numerical methods.
\end{abstract}

\end{frontmatter}


\section{Introduction}
\label{sec:intro}

Earthquakes are classically thought of as manifestations of frictional instabilities associated with the nucleation and subsequent rapid growth of slip on geologic fault surfaces.
The earthquake processes involve a wide range of complexities arising from the fault geometry, frictional contact, and off-fault damage (\ie~microcracks and temporal variations in the bulk wave speeds). 
These complexities are well known to have significant impacts on the fault rupture characteristics~\cite{harris1997effects,poliakov2002dynamic,dunham2003supershear,bhat2004dynamic,ma2015effect,ma2018strain,ma2019dynamic} and on the overall energy partitioning during the earthquake~\cite{sibson1977fault,andrews2005rupture,okubo2019dynamics}, which are critical to the analysis of natural and induced seismicity alike. 
Since most of these complexities are not tractable with analytical methods, numerical methods play an indispensable role in investigating source physics and its implications on seismological, geological, and geodetic observations.

For decades, the primary means for numerical modeling of fault rupture has been discontinuous approaches that treat the fault motion as a displacement jump (relative displacement) across a zero-thickness interface.
Notable examples include the boundary integral method~\cite{andrews1985dynamic,cochard1994dynamic,ben1997dynamic,kame1999simulation}, and the traction-at-split-node method which has been applied to finite difference~\cite{andrews1997wrinkle,andrews1999test,dalguer2007staggered} and finite element computations~\cite{duan2008effects,templeton2008off}. 
The popularity of these discontinuous approaches may be credited to the widespread applications of fracture mechanics in earthquake studies, in which the fault is considered a shear fracture with zero thickness.
This discontinuous viewpoint has further been justified by the fact that the fault slip is primarily localized in a region with a negligible thickness relative to both the fault length and the wavelengths of interest in the seismic wavefield~\cite{chester1986implications,chester1993internal,chester1998ultracataclasite,ben2003characterization}.  

For a discontinuous description of fault rupture, however, one should explicitly handle the geometric complexities involved in the seismogenesis---a particularly unsettling challenge when the fault path and/or the damage in the surrounding bulk would evolve.
For example, because the boundary integral method cannot model the material nonlinearity in surrounding rock masses, it is intrinsically unable to capture the growth of off-fault damage. 
The traction-at-split-node approach can simulate damage in rock masses by incorporating bulk inelasticity~\cite{andrews2005rupture,templeton2008off,dunham2011earthquake}, but it requires the fault interface to be aligned with the element boundaries.
Therefore, the traction-at-split-node method needs a sophisticated remeshing algorithm to model fault propagation in arbitrary directions. 
This limitation may be overcome by the use of embedded discontinuity methods such as the assumed enhanced strain (AES) method and the extended/generalized finite element method (XFEM/GFEM), which can accommodate the fault interface in the interior of discretized elements~\cite{borja2007continuum,foster2007embedded,liu2009extended,coon2011nitsche,liu2013extended}. 
Yet such embedded discontinuity methods may demand significant effort for implementation as they require enrichment of shape functions and sophisticated algorithms for numerical integration.

An alternative approach to the numerical modeling of fault rupture is to treat the fault surface as a continuous entity with finite thickness. 
Commonly used continuous approaches include the stress glut method~\cite{andrews1976rupture,andrews1999test,dalguer2006comparison} and thick fault zone methods~\cite{madariaga1998modeling,dalguer2006comparison,herrendorfer2018invariant,preuss2019seismic}. 
In contrast to the discontinuous approaches, these continuous approaches model the fault as an inelastic zone of finite thickness across a single or multiple layers of elements. 
The fault motion is then represented by an inelastic shear strain, which can be simply calculated according to a specific constitutive law for friction.
As a result, neither a remeshing algorithm nor explicit calculation of displacement jump is required for these continuous approaches, allowing them to be easily implemented in standard numerical methods.

However, the results of these continuous approaches are quite different from those produced by the discontinuous approaches.
Dalguer and Day~\cite{dalguer2006comparison} have shown that the results of the thick fault zone methods are not even qualitatively consistent with reference solutions obtained by standard discontinuous methods.
The results of the stress glut method show qualitative agreement with the reference solution, but they do not converge to the reference solution upon mesh refinement~\cite{dalguer2006comparison,preuss2019seismic}.
This non-convergence problem is attributed to the fact that these formulations lack a length scale that is necessary to retain mathematical well-posedness during softening behavior.
Such a convergence issue is absent in the very recent model proposed by Gabriel~\etal~\cite{gabriel2021unified}, which describes dynamic rupture processes in a low velocity fault zone using a regularized damage formulation.
Nevertheless, because the model is derived from damage rheology, there is no \textit{a priori} constraint on the results to converge to the laboratory-derived rate-and-state friction limits~\cite{dieterich1979modeling}. 
Also, while this approach is particularly promising, it entails several parameters that are not standard in earthquake studies and so would not be easy to calibrate.
Therefore, it is highly desirable to develop a continuous approach that consistently converges to the discontinuous fault description while relying on standard mechanics theory for shear fracture with frictional contact~\cite{palmer1973growth}.

Over the last decade, in the computational mechanics community, the phase-field method has emerged as a novel approach to continuous modeling of fracture.
The method diffusely approximates the sharp geometry of a crack surface with a spatially distributed variable---the phase field---and describes the evolution of the crack surface by a partial differential equation formulated from fracture mechanics theory.
In this way, the phase-field method handles complex crack geometries that are neither aligned with mesh boundaries nor do they require enrichment functions, lending itself to a simple implementation in standard numerical frameworks such as the finite element method. 
Furthermore, the phase-field method is free of the problems of the stress glut and thick fault methods.
Specifically, as the phase-field formulation is rooted in fracture mechanics theory, the phase-field solutions are consistent with those of discontinuous approaches based on the same theory. 
Also, because the phase-field approximation introduces a length scale for regularizing a sharp discontinuity, it is mathematically well-posed and hence its numerical solutions converge upon mesh refinement.
However, as the phase-field method was originally developed for tensile (mode \rn{1}) fracture, the vast majority of phase-field models in geomechanics have focused on tensile fracture, \eg~\cite{lee2016pressure,santillan2017phase,santillan2018phase,choo2018coupled,choo2018cracking,evans2020phase}. 

Recently, the capabilities of phase-field modeling have been extended to frictional shear fractures in geologic materials. 
Fei and Choo~\cite{fei2020phasea} have developed the first phase-field formulation for discontinuities with frictional contact.
Later, Fei and Choo~\cite{fei2020phaseb} have extended the formulation to propagating shear fractures, in a way that is demonstrably consistent with the fracture mechanics theory proposed by Palmer and Rice~\cite{palmer1973growth}. 
Also, Bryant and Sun~\cite{bryant2021phase} have incorporated rate- and state-dependent friction, which is considered essential to earthquake modeling, into the phase-field formulation for frictional discontinuities.

Yet none of the existing phase-field formulations is sufficiently credible for modeling fault rupture processes in earthquakes.
The current phase-field models for frictional shear fracture are all restricted to the quasi-static condition, neglecting dynamic effects or their quasi-dynamic approximation. 
The (quasi-)dynamic nature, however, is essential to the fault rupture modeling as it is responsible for several seismic features including fast rupture propagation and fault slip rates in the m/s range which correlates with strong ground motion. 
Also, a quasi-static model becomes unstable as the rupture accelerates and static equilibrium is no longer possible. 
It is thus important to explicitly retain the inertia term (a fully dynamic approach) or its approximation through radiation damping (a quasi-dynamic approach), to ensure the numerical stability of the calculation during the coseismic period, as well as to obtain results that are seismologically relevant. 

In this work, we develop the first phase-field approach to fault rupture and propagation.
First, we incorporate rate- and state-dependent friction into the most recently proposed phase-field framework for geologic discontinuities~\cite{fei2022phase}, whereby the displacement-jump-based kinematics of faults are transformed into a strain-based version and then inserted into the phase-field formulation for frictional interfaces. 
Second, we adopt, in this initial study, a quasi-dynamic formulation that approximates the inertial effects by a radiation damping term in the calculation of the shear stress on the fault plane~\cite{rice1993spatio}.
Compared with the fully-dynamic modeling, the quasi-dynamic formulation significantly reduces the computational cost, while retaining the numerical stability and qualitative patterns of the simulation results. 
Therefore, the quasi-dynamic formulation has been popular in numerical investigations of fault rupture processes, \eg~\cite{erickson2014efficient,thomas2014quasi,erickson2017finite,pampillon2018dynamic,abdelmeguid2019novel,erickson2020community,heimisson2020crack,cattania2021precursory,jiang2022community}.
Drawing on the same idea, we incorporate quasi-dynamic effects by augmenting a radiation damping term to the phase-field formulation for frictional shear fracture under quasi-static conditions~\cite{fei2020phaseb}. 
The resulting phase-field formulation is consistent with the earthquake energy budget and enables energy partitioning between strain energy, frictional work, and radiated energy approximated through the radiation damping. 
Through a number of numerical examples with varying complexity, we demonstrate the performance of the proposed formulation and its potential capabilities. 

At this point, we clarify two important limitations of the scope of this initial study.
First, as explained above, the formulation is limited to quasi-dynamic conditions.
Second, we shall limit the fault kinematics to a two-dimensional antiplane condition. 
Extension of the work to the fully dynamic limit and/or more complex kinematics will be presented in future work.

\section{Phase-field framework for frictional shear fracture}
\label{sec:pf-framework}

This section recapitulates a phase-field framework for general shear fracture with frictional contact, which has been developed in a series of recent works~\cite{fei2020phasea,fei2020phaseb,fei2021double,fei2022phase}.
The framework presented in this section will serve as the basis for the new phase-field formulation for geologic faults in the next section.

\subsection{Phase-field approximation and governing equations}
Consider the domain $\Omega$ delimited by boundary $\pd \Omega$. 
The boundary is decomposed into two parts, namely, the Dirichlet boundary, $\pd_{u} \Omega$, and the Neumann boundary, $\pd_{t} \Omega$, such that $\overline{\pd_{u} \Omega \cup \pd_{t} \Omega} = \pd \Omega$ and $\overline{\pd_{u} \Omega \cap \pd_{t} \Omega} = \emptyset$.
The domain may possess a set of discontinuities, which will be denoted by $\Gamma$. 
The time domain of interest is denoted by $\mathbb{T} := (0, t_{\max}]$. 

\begin{figure}[htbp]
  \centering
  \includegraphics[width=\textwidth]{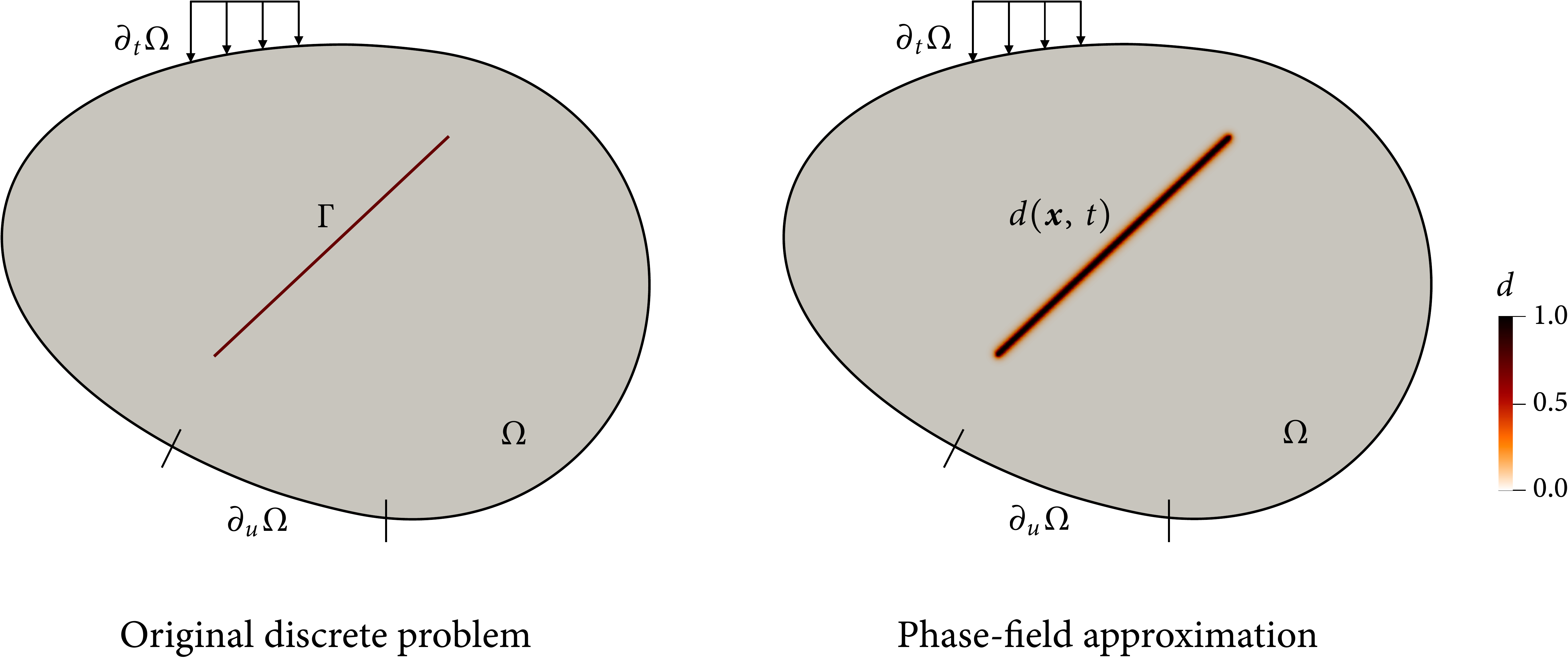}
  \caption{Phase-field approximation of sharp discontinuity. The sharp discontinuity $\Gamma$ on the left side is diffusely approximated by the phase-field variable $d$ on the right side.}
  \label{fig:pf-approx}
\end{figure}

Figure~\ref{fig:pf-approx} illustrates how the discontinuities $\Gamma$ can be diffusely approximated with the phase-field variable, $d$. 
The value of the phase-field variable ranges from 0 to 1, where $d=0$ denotes an intact (unfractured/undamaged) region and $d=1$ denotes a fully fractured/damaged region.  
As such, the phase-field variable can also be interpreted as the damage variable in damage mechanics~\cite{deborst2016gradient}.
The phase-field variable approximates the discontinuities through a crack density functional $\Gamma_d(d, \grad d)$ that satisfies
\begin{linenomath}
\begin{align}
	\int_{\Gamma} \: \dd A = \int_{\Omega} \delta_{\Gamma}(\tensor{x}) \: \dd V \approx \int_{\Omega} \Gamma_{d}(d, \grad d) \: \dd V , \label{eq:gamma-d-define}
\end{align}
\end{linenomath} 
where $\delta_{\Gamma}(\tensor{x})$ is the Dirac delta function. 
In this work, we adopt the crack density functional of the form
\begin{linenomath}
  \begin{align}
    \Gamma_{d}(d, \grad d) = \dfrac{3}{8} \left[\dfrac{d}{L} + L \grad d \cdot \grad d\right], \label{eq:crack-density-function-chosen}
  \end{align}
\end{linenomath}
where $L$ is a length parameter determining the width of the diffuse region.
The reason for choosing Eq.~\eqref{eq:crack-density-function-chosen}, instead of the most common crack density functional in the phase-field literature~\cite{miehe2010phase,borden2012phase}, is that it can lead to a family of phase-field formulations whose \revised{stress--strain responses} are insensitive to the length parameter $L$~\cite{geelen2019phase,fei2020phaseb}.
This length-insensitivity \revised{of the stress--strain response} is essential for modeling quasi-brittle materials such as rocks, for which the phase-field method should provide a prescribed peak strength and softening behavior regardless of the value of $L$.

Next, we introduce the two governing equations in the phase-field method. 
The first equation governs the displacement vector field, $\tensor{u}$, based on the linear momentum balance (neglecting the body force) 
\begin{linenomath}
\begin{align}
	\div \tstress(\tensor{u},d) = \rho \ddot{\tensor{u}} \quad \text{in} \:\: \Omega \times \mathbb{T} ,  \label{eq:momentum-balance-full-dynamic}
\end{align}
\end{linenomath}
where $\tstress$ is the stress tensor, $\rho$ is the mass density, and $\ddot{\tensor{u}}$ is the acceleration vector at the local material point.
It is noted that the inertial term $\rho\ddot{\tensor{u}}$ can be neglected under quasi-static conditions, but it should be retained or approximated otherwise.
Also, as explained earlier, we shall restrict our attention to two-dimensional antiplane problems. 
In this case, only the displacement in the $z$ direction remains, and the momentum balance equation reduces to 
\begin{linenomath}
\begin{align}
   \dfrac{\pd \stress_{xz}(u_z,d)}{\pd x} + \dfrac{\pd \stress_{yz}(u_z,d)}{\pd y} = \rho \ddot{u}_{z}  \quad \text{in} \:\: \Omega \times \mathbb{T} . 
\end{align}
\end{linenomath}
For notational convenience, we define a vector representing the two shear stress components, as
\begin{linenomath}
\begin{align}
  \tensor{\tau} &:= 
  \begin{bmatrix}
    \stress_{xz} & \stress_{yz}
  \end{bmatrix}^{\intercal} . 
\end{align}
\end{linenomath}
Now the momentum balance equation can be rewritten as 
\begin{linenomath}
\begin{align}
   \div \tensor{\tau} = \rho \ddot{u}_{z} \quad \text{in} \:\: \Omega \times \mathbb{T} ,  \label{eq:momentum-balance-antiplane}
\end{align}
\end{linenomath}
Accordingly, the boundary conditions can be written as
\begin{linenomath}
\begin{align}
   \tensor{v} \cdot \tensor{\tau} = \hat{t}_{z} \quad &\text{on} \:\: \pd_{t} \Omega \times \mathbb{T},  \\
   u_{z} = \hat{u}_{z} \quad &\text{on} \:\: \pd_{u} \Omega \times \mathbb{T},
\end{align}
\end{linenomath}
where $\hat{t}_{z}$ and $\hat{u}_{z}$ are the prescribed shear traction and displacement on their corresponding boundaries, and $\tensor{v}$ denotes the unit vector normal to the boundary. 

The second governing equation describes the evolution of the phase field, $d$, based on fracture mechanics theory. 
Following the derivation in Fei and Choo~\cite{fei2020phaseb}, the phase-field evolution equation can be obtained as 
\begin{linenomath}
\begin{align}
  -g'(d) \mathcal{H}^{+}(u_{z}) + \dfrac{3\mathcal{G}_{\rn{2}}}{8L}(2L^2 \div \grad d - 1)= 0 \quad \text{in} \:\: \Omega \times \mathbb{T}. \label{eq:pf-evolution} 
\end{align}
\end{linenomath} 
Here, $\mathcal{H}^{+}$ is the effective crack driving force, $\mathcal{G}_{\rn{2}}$ is the critical fracture energy for shear crack, and $g(d)$ is the degradation function which accounts for the degradation of material stiffness due to fracturing.
This phase-field evolution equation describes a balance between the dissipation of stored energy (\ie~$-g'(d)\mathcal{H}^{+}$) and the generation of new surface energy (\ie~$3 \mathcal{G}_{\rn{2}}/8L (2L^2 \div \grad d - 1))$ during the fracturing process, which is consistent with the premise of fracture mechanics after Griffith~\cite{griffith1921phenomena}.  
The specific expression of $g(d)$ that pairs with the adopted crack density functional~\eqref{eq:crack-density-function-chosen} is given by
\begin{linenomath}
\begin{align}
  g(d) = \dfrac{(1 - d)^{n}}{(1 - d)^{n} + md(1 + pd)} \quad \text{with} \:\: m = \dfrac{3\mathcal{G}_{\rn{2}}}{8L} \dfrac{1}{\mathcal{H}_{t}} . \label{eq:gd} 
\end{align}
\end{linenomath}
Here, $n$ and $p$ are modeling parameters, for which we adopt $n = 2$ and $p=20$ in this work.
It is noted that the specific combination of Eqs.~\eqref{eq:crack-density-function-chosen} and~\eqref{eq:gd} is essential to attain the length-insensitivity of the phase-field model, see Fei and Choo~\cite{fei2020phaseb} for details.
The remaining term $\mathcal{H}_{t}$ in Eq.~\eqref{eq:gd} represents a threshold value of the crack driving force at the material's peak strength.
For frictional shear fracture, $\mathcal{H}_{t}$ is derived as~\cite{fei2020phaseb}
\begin{linenomath}
\begin{align}
  \mathcal{H}_{t} = \dfrac{1}{2G} (\tau_{p} - \tau_{r})^2 ,  \label{eq:H-threshold}
\end{align}
\end{linenomath}
where $G$ is the shear modulus of the bulk material, and $\tau_{p}$ and $\tau_{r}$ denote the peak and residual shear strengths, respectively.
For cohesive--frictional materials, $\tau_{p}$ and $\tau_{r}$ are usually related to the normal pressure on the fracture, $p_{\cn}$, through a friction coefficient.
For simplicity, let us assume that $\tau_{p}$ and $\tau_{r}$ share the same friction coefficient.
Then the peak and residual strengths can be written as
\begin{linenomath}
\begin{align}
  \tau_{p} &= c + p_{\cn} \mu , \\
  \tau_{r} &=  p_{\cn} \mu,  
  \label{eq:tau-r}
\end{align}
\end{linenomath}
where $c$ denotes the cohesive strength, and $\mu$ denote the friction coefficient, respectively. 
Also, it is noted that while $\mu$ was assumed to be constant in the existing phase-field formulation for frictional shear fracture~\cite{fei2020phaseb}, the assumption is relaxed here to incorporate rate- and state-dependent friction, as will be explained shortly.

\subsection{Kinematics of phase-field fracture}
To close the governing equations~\eqref{eq:momentum-balance-antiplane} and \eqref{eq:pf-evolution}, constitutive laws for the bulk material and the fracture region should be introduced. 
While an existing constitutive law for the bulk material---based on the strain field---may be directly incorporated into the phase-field setting, that for a fracture surface---based on the displacement or velocity jump---is incompatible with the phase-field setting.
Therefore, to utilize an advanced constitutive law for frictional discontinuities such as a rate- and state-dependent friction law, the displacement-jump-based kinematics of an existing constitutive law should be transformed into its strain-based counterpart.
In what follows, we describe how to perform such transformation, specializing the strategy developed in Fei~\etal~\cite{fei2022phase} to two-dimensional antiplane problems. 

Starting from the standard displacement-jump-based kinematics, we decompose the total antiplane displacement, $u_{z}$, into a component denoting the solid deformation, $\bar{u}_{z}$, and a component denoting the displacement jump, $[\![ u_{z} ]\!]$, as
\begin{linenomath}
\begin{align}
    u_{z} = \bar{u}_{z} + [\![ u_{z} ]\!] H_{\Gamma} (\tensor{x}). \label{eq:disp-decompose}
\end{align}
\end{linenomath}
Here, $H_{\Gamma}$ is the Heaviside function, defined as 
\begin{linenomath}
\begin{align}
    H_{\Gamma} (\tensor{x}) := 
    \left \{
    \begin{array}{ll}
         1 & \text{if} \:\: \tensor{x} \in \Omega_{+},   \\
         0 & \text{if} \:\: \tensor{x} \in \Omega_{-}. 
    \end{array}
    \right .
\end{align}
\end{linenomath}
In this two-dimensional antiplane setup, the displacement jump may be rewritten as
\begin{linenomath}
\begin{align}
    [\![ u_{z} ]\!] = \alpha \zeta_{f}  , \label{eq:disp-jump}
\end{align}
\end{linenomath}
where $\zeta_{f}$ denotes the magnitude of the antiplane slip, and $\alpha = \pm 1$ denotes the slip direction. 
Specifically, if the slip direction is in the positive $z$ direction, then $\alpha = 1$; otherwise, $\alpha = -1$. 
Then, inserting Eq.~\eqref{eq:disp-jump} into Eq.~\eqref{eq:disp-decompose} and taking gradients on both sides of Eq.~\eqref{eq:disp-decompose}, we can express the shear strain as
\begin{linenomath}
\begin{align}
    \tensor{\gamma} := \grad u_{z} = \bar{\tensor{\gamma}} + H_{\Gamma}(\tensor{x}) \alpha \grad \zeta_{f} + \alpha \zeta_{f} \tensor{n} \delta_{\Gamma}(\tensor{x}) . \label{eq:strain-decompose}
\end{align}
\end{linenomath}
Here, $\tensor{\gamma}$ represents the vector form of the shear strains in two-dimensional antiplane problems, given by 
\begin{linenomath}
\begin{align}
  \tensor{\gamma} &= 
  \begin{bmatrix}
    \dfrac{\pd u_{z}}{\pd x} & \dfrac{\pd u_{z}}{\pd y}
  \end{bmatrix}^{\intercal} . 
\end{align}
\end{linenomath}
Also, $\bar{\tensor{\gamma}} := \grad \bar{u}_{z}$ denotes the continuous part of the shear strain vector, $\tensor{n}$ denotes the unit vector normal to the discontinuity, and the Dirac delta function $\delta_{\Gamma}(\tensor{x})$ comes from the gradient of the Heaviside function, \ie~$\grad H_{\Gamma}(\tensor{x}) = \tensor{n} \delta_{\Gamma}(\tensor{x})$. 
If we assume that the slip does not vary along the discontinuity, \ie~$\grad \zeta_{f} = \tensor{0}$, Eq.~\eqref{eq:strain-decompose} reduces to 
\begin{linenomath}
\begin{align}
    \tensor{\gamma} = \bar{\tensor{\gamma}} + \alpha \zeta_{f} \tensor{n} \delta_{\Gamma}(\tensor{x}) . \label{eq:strain-decompose-simple}
\end{align}
\end{linenomath}
Lastly, by appealing to Eq.~\eqref{eq:gamma-d-define}, we approximate the Dirac delta function in Eq.~\eqref{eq:strain-decompose-simple} by the crack density functional and obtain
\begin{linenomath}
\begin{align}
   \tensor{\gamma} \approx \bar{\tensor{\gamma}} + \alpha \zeta_{f} \tensor{n} \Gamma_{d}(d, \, \grad d) . \label{eq:strain-decompose-pf} 
\end{align}
\end{linenomath}
The above equation transforms the discontinuous kinematics into a strain-based continuous description in a way consistent with the phase-field method.

\subsection{Constitutive relations for phase-field fracture}
Based on the kinematic description derived above, we now introduce constitutive laws following the phase-field formulation for interfaces with frictional contact~\cite{fei2020phasea}.
The phase-field formulation decomposes the shear stress tensor into the shear stress tensor in the bulk matrix, $\tensor{\tau}_{m}$, and that in the fracture, $\tensor{\tau}_{f}$, weighting them according to the value of $g(d)$.
In a two-dimensional antiplane setting, the stress decomposition can be written as
\begin{linenomath}
\begin{align}
    \tensor{\tau} = g(d) \tensor{\tau}_{m} + [1 - g(d)] \tensor{\tau}_{f} , \label{eq:stress-decompose} 
\end{align}
\end{linenomath}
In this work, we shall assume that the bulk material is linear elastic.
The constitutive relation for the matrix is then given by
\begin{linenomath}
\begin{align}
  \tensor{\tau}_{m} = G \tensor{\gamma},
\end{align}
\end{linenomath}
Likewise, the constitutive relation for the fracture can be introduced by relating $\tensor{\tau}_{f}$ with the strain component accounting for continuous deformation, $\bar{\tensor{\gamma}}$~\cite{regueiro2001plane}. It gives
\begin{linenomath}
\begin{align}
  \tensor{\tau}_{f} = G \bar{\tensor{\gamma}} = G \left[ \tensor{\gamma} - \alpha \zeta_{f} \tensor{n} \Gamma_{d}(d, \grad d) \right] .  \label{eq:stress-fracture-general}
\end{align}
\end{linenomath} 
Note that the value of $\tensor{\tau}_{f}$ is constrained frictional sliding.
The constraint can be expressed by a yield function of the following form
\begin{linenomath}
\begin{align}
  F(\tensor{\tau}_{f}) = \lvert \tau_{f} \rvert - \mu p_{\cn}
   \leq 0 , \label{eq:yield-function}
\end{align}  
\end{linenomath}
where $\tau_{f} := \tensor{\tau}_{f} \cdot \tensor{n}$ is the resolved shear stress on the discontinuity.
The discontinuity is in the stick mode when $F<0$ and in the slip mode when $F=0$.

\section{Phase-field formulation for quasi-dynamic faulting}
\label{sec:formulation}

In this section, we develop a phase-field formulation for quasi-dynamic fault nucleation, growth, and propagation, specializing and extending the general framework described in the previous section. 
Specifically, here we incorporate rate- and state-dependent friction, radiation damping, and their impacts on the faulting process.

\subsection{Rate- and state-dependent friction}
To incorporate rate- and state-dependent \revised{friction}, we shall let the friction coefficient, $\mu$, in Eq.~\eqref{eq:yield-function} be a function of the slip rate and the state variable of the fault.
This is commonly described by the Dieterich--Ruina friction law~\cite{dieterich1979modeling,ruina1983slip}, given by
\begin{linenomath}
\begin{align}
  \mu(V, \theta) := \mu_0 + a \ln \left(\dfrac{V}{V_{0}} \right) + b \ln \left( \dfrac{\theta V_{0}}{D_{c}} \right).  
  \label{eq:rs-friction-original}
\end{align}
\end{linenomath}
Here, $\mu_{0}$ is the reference friction coefficient, $V := \dot{\zeta}_{f}$ is the slip rate,$a$ and $b$ are constitutive parameters that account for the rate-and-state dependence, $V_{0}$ is the reference slip rate, $D_{c}$ is the characteristic slip distance for state variable evolution, and $\theta$ is the state variable \revised{which corresponds to the average lifetime of asperity contact and thus has the unit of time.
It is noted that the state variable should be non-negative by definition.
Also, the state variable can evolve,} and the evolution is assumed to follow the aging law proposed by Rice and Ruina~\cite{rice1983stability}, namely,
\begin{linenomath}
\begin{align}
	\dot{\theta} = 1 - \dfrac{V\theta}{ D_{c}} . \label{eq:aging-law}
\end{align}
\end{linenomath}
\revised{According to the aging law, the state variable increases ($\dot{\theta} > 0$) when the fault is stick ($V=0$), and its evolution rate becomes slower as the fault begins to slip. When the slip rate becomes $D_{c}/\theta$, the state reaches a steady state ($\dot{\theta} = 0$).}
The rate- and state-dependent friction coefficient at this steady state value is 
\begin{linenomath}
\begin{align}
  \mu_{ss} = \mu_{0} + (a - b) \ln \left(\dfrac{V}{V_{0}} \right) .
\end{align}
\end{linenomath}
The above equation indicates that, if $a > b$, then the fault has velocity-strengthening friction, as the steady-state friction increases with the slip rate. 
Otherwise, if $a < b$, the fault has velocity-weakening friction, which may lead to seismic instabilities.

Notably, the original rate- and state-dependent friction law~\eqref{eq:rs-friction-original} has a singularity at $V = 0$.
To avoid this singularity in numerical modeling, we adopt a regularized rate- and state-dependent friction~\cite{rice1996slip}, given by
\begin{linenomath}
\begin{align}
  \mu(V, \theta) := a \sinh^{-1}\left[\dfrac{V}{2V_{0}} \exp \left(\dfrac{\mu_{0} + b \ln (V_{0}\theta / D_{c})}{a} \right) \right] .  \label{eq:rs-friction-regularize}
\end{align}
\end{linenomath} 

\subsection{Quasi-dynamic approximation}

As an initial attempt for phase-field modeling of fault rupture, this work adopts a quasi-dynamic approach proposed by Rice~\cite{rice1993spatio} to accommodate inertial effects.
This approach approximates the inertial effects by a radiation damping term, which is augmented to the calculation of the shear stress on the fault surface.
Apart from its simplicity, this quasi-dynamic approximation stabilizes accelerated fault slip in the absence of inertia effects~\cite{rice1993spatio}. 
Thanks to this simplicity and efficacy, the quasi-dynamic approach has been widely used in numerical simulations of fault rupture, see, \eg~\cite{erickson2014efficient,thomas2014quasi,erickson2017finite,pampillon2018dynamic,abdelmeguid2019novel,erickson2020community,heimisson2020crack,cattania2021precursory,jiang2022community}.

The radiation damping term can be calculated as $\eta V$, where $\eta$ is a damping coefficient given by~\cite{rice1993spatio}
\begin{linenomath}
\begin{align}
  \eta = \dfrac{G}{2c_{s}}, 
\end{align}
\end{linenomath}
with $c_{s}$ denoting the shear wave speed. 
In the quasi-dynamic formulation, the radiation damping term plays the role of an additional shear strength contribution to the rate- and state frictional fault strength.
Therefore, for the phase-field formulation at hand, the radiation damping term is added to the yield function~\eqref{eq:yield-function} as
\begin{linenomath}
  \begin{align}
    F(\tensor{\tau}_{f}) = \lvert \tau_{f} \rvert - p_{\cn}\mu(V, \theta) - \eta V \leq 0. \label{eq:yield-function-quasidynamic}
  \end{align}
\end{linenomath}

\subsection{Crack driving force for quasi-dynamic fault propagation}
In the phase-field formulation, the crack driving force can be derived as (see Fei and Choo~\cite{fei2020phaseb} for details)
\begin{linenomath}
  \begin{align}
    g'(d) \mathcal{H}^{+} = \dfrac{\pd\psi}{\pd d}.  
  \end{align}
\end{linenomath}
where $\psi$ is the potential energy density. 
As shown above, the crack driving force is calculated as the rate of energy change that balances the energy dissipation from fracture generation.
In the existing phase-field model for quasi-static shear fracture~\cite{fei2020phaseb}, the strain energy and the frictional energy dissipation are considered as in the fracture mechanics theory of Palmer and Rice~\cite{palmer1973growth}.
For (quasi-)dynamic shear fracture, however, the radiated energy (approximated by radiation damping dissipation) should also be considered, to be consistent with the earthquake energy budget presented in Kanamori~\etal~\cite{kanamori2000microscopic} and Abercrombie and Rice~\cite{abercrombie2005can}.

Therefore, here we calculate $\psi$ as
\begin{linenomath}
  \begin{align}
    \psi = \psi^{\el} + \psi^{\mathrm{f}} + \psi^{\mathrm{r}} \label{eq:psi-dynamic} 
  \end{align}
\end{linenomath}
where $\psi^{\el}$ denotes the strain energy density, $\psi^\mathrm{f}$ the friction dissipation density, and
$\psi^\mathrm{r}$ the viscous dissipation density emanating from the radiation damping term in the quasi-dynamic formulation.
This viscous dissipation term, which is newly introduced in the current formulation, can be viewed as an approximation of the radiated energy due to seismic wave propagation. 

The three energy densities in Eq.~\eqref{eq:psi-dynamic} are expressed in rate forms as they are contact-dependent and hence incrementally nonlinear. 
The specific expressions are given as follows. 
First, because the densities of the strain energy and friction dissipation have been derived in Fei and Choo~\cite{fei2020phaseb}, here we only need to reformulate them in the two-dimensional antiplane setting. 
The rate of the strain energy density is expressed as 
\begin{linenomath}
\begin{align}
  \dot{\psi}^{\el} = \left \{
  \begin{array}{ll}
    \tensor{\tau}_{m} \cdot \dot{\tensor{\gamma}} & \text{if}\:\:\text{stick,} \\
    \tensor{\tau}_{m} \cdot \dot{\tensor{\gamma}} - \left[1 - g(d)\right] \tau_{m} \dot{\gamma} & \text{if}\:\:\text{slip,}
  \end{array}
  \right . \label{eq:strain-energy-density-rate}
\end{align}
\end{linenomath}
where $\tau_{m} := \tensor{\tau}_{m} \cdot \tensor{n}$, and $\gamma := \tensor{\gamma}\cdot \tensor{n}$.  
The rate of the friction dissipation density is formulated as 
\begin{linenomath}
\begin{align}
  \dot{\psi}^{\mathrm{f}} = \left \{
  \begin{array}{ll}
    0 & \text{if stick,} \\
    \left[1 - g(d)\right] \tau_{r} \dot{\gamma} & \text{if}\:\:\text{slip.}
  \end{array}
  \right .  \label{eq:friction-dissipation-density-rate}
\end{align}
\end{linenomath}
For the rate of the viscous dissipation density, which is a new term in this work, recall that the radiation damping term $\eta V$ can be considered as an additional shear strength to the frictional strength, $\tau_{r}$ (\cf~Eq.~\eqref{eq:yield-function}). 
Therefore, the viscous dissipation density can be formulated in the same way as the friction dissipation.
The rate form of the viscous dissipation is thus given by 
\begin{linenomath}
\begin{align}
  \dot{\psi}^{\mathrm{r}} = \left \{
  \begin{array}{ll}
    0 & \text{if stick,} \\
    \left[1 - g(d)\right] V \eta \dot{\gamma} & \text{if}\:\:\text{slip.}
  \end{array}
  \right .  \label{eq:viscous-dissipation-density-rate}
\end{align}
\end{linenomath}
It is noted that both the friction and viscous dissipation terms remain zero under a stick condition. 

Next, we integrate the three energy density rates in Eqs.~\eqref{eq:strain-energy-density-rate}, \eqref{eq:friction-dissipation-density-rate} and \eqref{eq:viscous-dissipation-density-rate} in time and differentiate the resultant expressions with respect to $d$. 
These integration and differentiation can be done through the procedure in Fei and Choo~\cite{fei2020phaseb}, so their details are omitted for brevity. 
Eventually, the crack driving force is obtained as
\begin{linenomath}
\begin{align}
  \mathcal{H}^{+}= \left \{
  \begin{array}{ll}
    \max_{t \in [0, t_{\max}]} \mathcal{H}^{+}(t) & \text{if}\:\:\text{stick,} \\
    \mathcal{H}_{t} + \mathcal{H}_{\text{slip}} & \text{if}\:\:\text{slip,}
  \end{array}
  \right . \label{eq:psi-dynamic-stick-slip}
\end{align}
\end{linenomath}
where 
\begin{linenomath}
\begin{align}
  \mathcal{H}_\text{slip} := \int_{\gamma_{p}}^{\gamma} (\tau_{m} - \tau_{r} - \eta V)\: \dd \gamma. 
\end{align}
\end{linenomath}
Here, $\mathcal{H}_\text{slip}$ represents the crack driving force contributed by the quasi-dynamic frictional slip after the material reaches its peak strength, and $\gamma_{p}$ denotes the shear strain in the slip direction when $\tau_{m} = \tau_{p}$. 
Also, in Eq.~\eqref{eq:psi-dynamic-stick-slip}, the crack driving force at the stick state is evaluated by its history maximum to ensure the irreversibility of fault propagation~\cite{fei2020phaseb}. 
Further, $\mathcal{H}^{+}$ is set to be equal to $\mathcal{H}_{t}$ to ensure no damage ($d=0$) before the intact material reaches its peak strength.
So the final formulation of the crack driving force can be written as
\begin{linenomath}
\begin{align}
  \mathcal{H}^{+}= \left \{
  \begin{array}{ll}
    \mathcal{H}_{t} & \text{if}\:\:\text{intact,} \\
    \max_{t \in [0, t_{\max}]} \mathcal{H}^{+}(t) & \text{if}\:\:\text{stick,} \\
    \mathcal{H}_{t} + \mathcal{H}_{\text{slip}} & \text{if}\:\:\text{slip.}
  \end{array}
  \right . \label{eq:psi-dynamic-final}
\end{align}
\end{linenomath}

\subsection{Determination of the fault propagation direction}
\label{sec:direction}

The final task to complete the formulation is to determine the direction of fault propagation---equivalently, the unit normal vector $\tensor{n}$ in the calculation of the stress and crack driving force. 
For this purpose, it is postulated that the fault propagates in the direction that maximizes the crack driving force, as assumed in the existing phase-field models for quasi-static fractures \cite{bryant2018mixed,fei2020phaseb,fei2021double}.
According to Eq.~\eqref{eq:psi-dynamic-final}, the increment of the crack driving force is proportional to $(\tau_{m} - \tau_{r} - \eta V)$. 
When the fault starts to grow, $V=0$, and thus the radiation damping term in the crack driving force is zero.
Also, since the normal stress is prescribed in this two-dimensional antiplane setting, the frictional strength $\tau_{r}$ is unaffected by the propagation direction. 
Therefore, the crack driving force is maximized when $\tau_{m}$ is maximum. 
Now, to search for the direction that maximizes $\tau_{m}$, let us denote by $\vartheta$ the angle between the $x$ axis and the potential fault plane. 
Then the unit normal vector can be written as 
\begin{linenomath}
\begin{align}
  \tensor{n} = 
  \begin{bmatrix}
     \sin \vartheta & \cos \vartheta 
  \end{bmatrix}^{\intercal} . 
\end{align}
\end{linenomath}
Thus, $\tau_{m}$ can be calculated as
\begin{linenomath}
\begin{align}
  \tau_{m} (\vartheta) = \tensor{\tau}_{m} \cdot \tensor{n} = \stress_{m,xz} \sin \vartheta + \stress_{m,yz} \cos \vartheta . 
\end{align}
\end{linenomath}
Since the value of $\tau_{m}$ is maximum when $\tau'_{m}(\vartheta) = 0$, we get
\begin{linenomath}
\begin{align}
  \vartheta = \arctan\left( \dfrac{\stress_{m, xz}}{\stress_{m, yz}} \right) . 
\end{align}
\end{linenomath}
Accordingly, the unit normal vector $\tensor{n}$ is calculated as
\begin{linenomath}
\begin{align}
  \tensor{n} = 
  \begin{bmatrix} 
     \dfrac{\stress_{m,xz}}{\lvert \tensor{\tau}_{m} \rvert} & \dfrac{\stress_{m,yz}}{\lvert \tensor{\tau}_{m} \rvert }
  \end{bmatrix}^{\intercal} , 
\end{align}
\end{linenomath}
where $\lvert \tensor{\tau}_{m} \rvert := \sqrt{\stress^{2}_{m,xz} + \stress^{2}_{m,yz}}$ denotes the magnitude of the shear stress vector in the intact material. 

\subsection{Summary of equations}
Before closing this section, we summarize the key equations of the proposed phase-field formulation. 
The two governing equations under quasi-dynamic, two-dimensional anti-plane conditions are 
\begin{linenomath}
\begin{align}
	\div \tensor{\tau} (u_{z}, d) &= 0 \quad \text{in} \:\: \Omega \times \mathbb{T},   \label{eq:momentum-balance-quasidynamic-antiplane} \\ 
	-g'(d) \mathcal{H}^{+} + \dfrac{3 \mathcal{G}_{\rn{2}}}{8L}(2 L^2 \div \grad d - 1) &= 0 \quad \text{in} \:\: \Omega \times \mathbb{T}. \label{eq:pf-evolution-quasidynamic-antiplane} 
\end{align}
\end{linenomath}
Here, the stress $\tensor{\tau}$ and the crack driving force $\mathcal{H}^{+}$ are evaluated differently in three different states, namely (i) when the material is intact (unfractured), (ii) when the material is fractured ($d>0$) and in a stick mode, and (iii) when the material is fractured ($d>0$) and in a slip mode. 
The equations of the stress and the crack driving force in these three states are
\begin{linenomath}
\begin{align}
	\left . 
		\begin{array}{r}
			\dot{\tensor{\tau}} = G \dot{\tensor{\gamma}} \\ 
			\quad \mathcal{H}^{+} = \mathcal{H}_{t}
		\end{array}
	\right \}
	\text{if}\:\:\text{intact},
\end{align}
\end{linenomath}
\begin{linenomath}
  \begin{align}
	\left . 
		\begin{array}{r}
			\dot{\tensor{\tau}} = G \dot{\tensor{\gamma}} \\ 
			\quad \mathcal{H}^{+} = \max_{t \in [0, t_{\max}]} \mathcal{H}^{+} (t)
		\end{array}
	\right \}
	\text{if}\:\:\text{stick},
\end{align}
\end{linenomath}
and 
\begin{linenomath}
\begin{align}
	\left . 
		\begin{array}{r}
			\dot{\tensor{\tau}} = G \dot{\tensor{\gamma}} - [ 1- g(d)] \alpha \dot{\zeta}_{f} \tensor{n} \Gamma_d (d, \grad d) \\ 
			\quad \mathcal{H}^{+} = \mathcal{H}_{t} + \int_{\gamma_{p}}^{\gamma} (\tau_{m} - \tau_{r} -\eta V) \: \dd \gamma
		\end{array}
	\right \}
	\text{if}\:\:\text{slip}.
\end{align}
\end{linenomath}
Note that here the stress is given in rate form because of the incremental nonlinearity and history dependence of the stress evolution. 
It is noted that the rate of the slip magnitude $\dot{\zeta}_{f}$ in the slip state is calculated by solving $F = 0$, where $F$ denotes the yield function in Eq.~\eqref{eq:yield-function-quasidynamic}. 

The governing equations Eqs.~\eqref{eq:momentum-balance-quasidynamic-antiplane} and \eqref{eq:pf-evolution-quasidynamic-antiplane} can be well solved by the standard finite element method.
Descriptions of the finite element discretization and the key solution algorithms are provided in~\ref{appendix:discretization}.

\section{Numerical examples}
\label{sec:numeric}

The purpose of this section is twofold: (i) to verify the proposed phase-field model, 
and (ii) to demonstrate the capability of the phase-field model for simulating quasi-dynamic faulting processes under complex geometric and material conditions. 
For the first purpose, we design a benchmark example of a non-propagating fault and compare the phase-field solutions to this example with solutions obtained by a well-verified discontinuous approach.
For the second purpose, we model fault propagation and damage growth under more complex conditions including material heterogeneity. 

For all numerical examples, we use the same set of material parameters adopted from a SCEC/USGS verification example TPV102~\cite{harris2009scec,harris2018suite}, which are  presented in Table~\ref{tab:fault-parameters}.
It is noted that the parameters give the shear modulus of rock masses as $G = \rho c_{s}^{2} = 32.04$ GPa. 
Also noted is that because $a<b$, the fault of interest has velocity-weakening friction.

\begin{table}[htbp]
    \centering
    \begin{tabular}{l|l|l|c}
    \toprule
    Parameter & Symbol & Unit & Value  \\
    \midrule
    Mass density & $\rho$ & $\text{kg/m}^{3}$ & 2670 \\
    Shear wave speed & $c_{s}$ & km/s & 3.464 \\
    Characteristic slip distance & $D_{c}$ & m & 0.02 \\ 
    Reference slip rate & $V_{0}$ & m/s & $10^{-6}$ \\ 
    Initial reference friction coefficient & $\mu_{0, \text{init}}$ & -- & 0.6 \\ 
    Minimum reference friction coefficient after nucleation & $\mu_{0, \min}$ & -- & 0.4 \\ 
    Rate-and-state parameter for the direct effect & $a$ & -- & 0.008 \\ 
    Rate-and-state parameter for state evolution & $b$ & -- & 0.012 \\ 
    \bottomrule
    \end{tabular}
    \caption{Material parameters for the fault in the numerical simulations.}
    \label{tab:fault-parameters}
\end{table}

In all the examples, we assign a constant normal pressure $p_{\cn} = 50$ MPa on the fault and set the initial slip rate as $V_\text{init} =10^{-9}$ m/s. 
Inserting this initial slip rate into Eq.~\eqref{eq:rs-friction-regularize} and assuming that the fault is initially in a steady state (\ie~$\dot{\theta} = 0$), we get the initial friction coefficient as
\begin{linenomath}
\begin{align}
  \mu_\text{init} = a \sinh^{-1}\left[\dfrac{V_\text{init}}{2V_{0}} \exp \left(\dfrac{\mu_{0, \text{init}} + b \ln(V_{0}/V_\text{init})}{a} \right)\right] \approx 0.628 .
\end{align}
\end{linenomath}
Then, by solving $F=0$ in Eq.~\eqref{eq:yield-function-quasidynamic}, we get the initial shear stress on the fault as $\tau_{f,\text{init}} \approx 31.382$ MPa.

All the phase-field results presented in this section are produced by an in-house finite element solver built on the \texttt{deal.II} finite element library~\cite{arndt2019deal,arndt2021deal}.
\revised{We use structured meshes comprised of linear quadrilateral elements and standard Gauss quadrature.}
The reference result for the verification example is produced by a well-verified discontinuous method called the hybrid finite element and the spectral boundary integral (FEBE) method~\cite{ma2019hybrid,abdelmeguid2019novel}.
\revised{The FEBE method models a fault as a sharp interface inside a predefined fault zone of finite thickness, and discretizes the fault zone with finite elements so that the method can incorporate geometric and material complexities in the fault zone.
Meanwhile, the FEBE method models the rest of the domain as a homogeneous and linear elastic material, and truncates it away with the spectral boundary integral equation for computational efficiency.} 
In this comparison exercise, the spectral boundary integral equation part in FEBE is turned off and only the finite element part is used to discretize the bounded domain.

\subsection{Rupture of a fault}
\begin{figure}[htbp]
  \centering
  \includegraphics[width=0.6\textwidth]{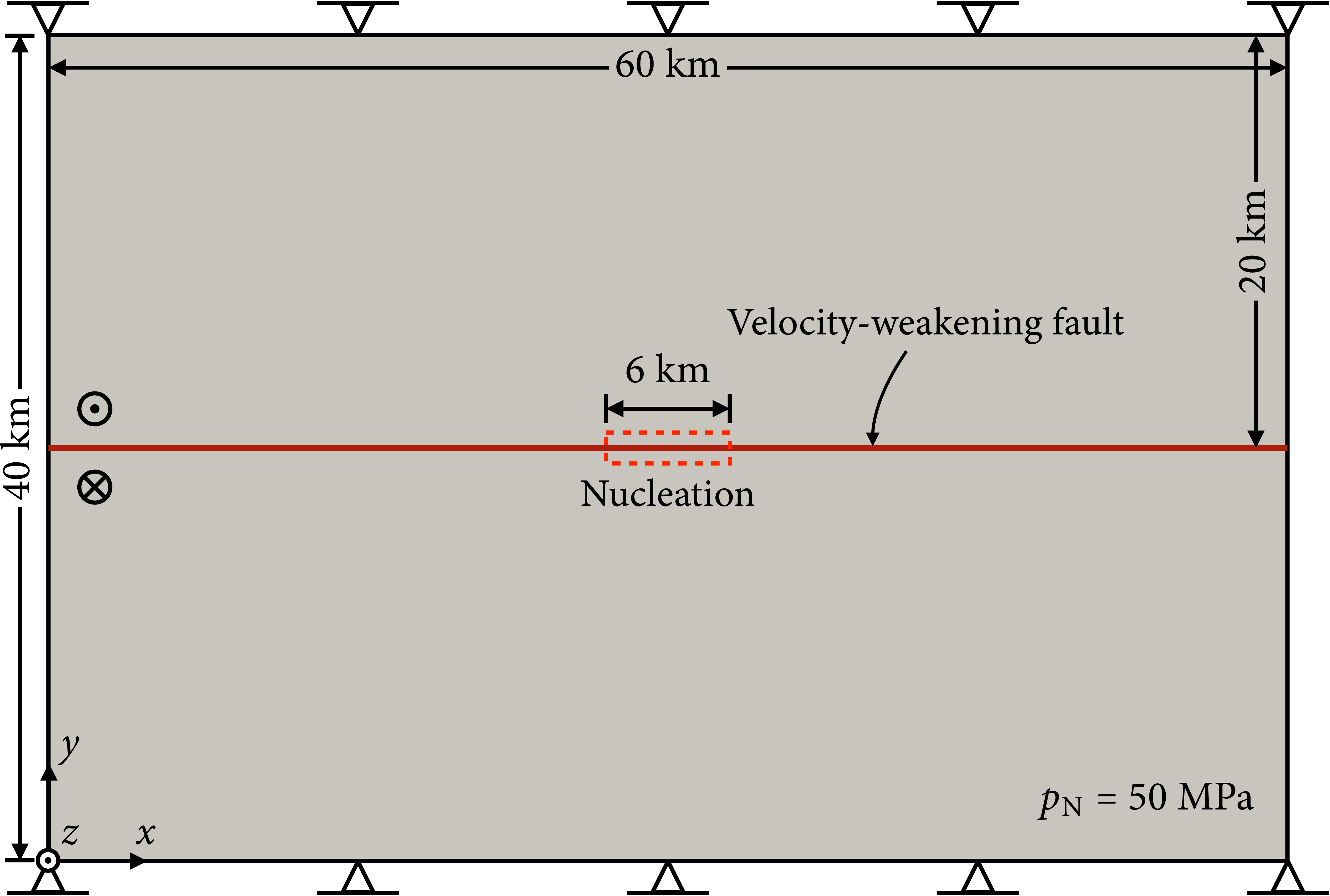}
  \caption{Rupture of a fault: geometry and boundary conditions.}
  \label{fig:fault-rupture-setup}
\end{figure}

The first example is intended to verify the phase-field approach by comparing it with a discontinuous approach that has already been well verified in the literature.
For this purpose, we design a quasi-dynamic fault problem, of which the domain geometry and boundary conditions are depicted in Fig.~\ref{fig:fault-rupture-setup}.
The domain is 60 km wide and 40 km high, being intersected by a horizontal fault located in the middle of the domain ($y = 20$ km). 
A 6 km wide nucleation region is assigned at the center of the fault. The details of the nucleation procedure will be described later.

\begin{figure}[htbp]
  \centering
  \includegraphics[width=\textwidth]{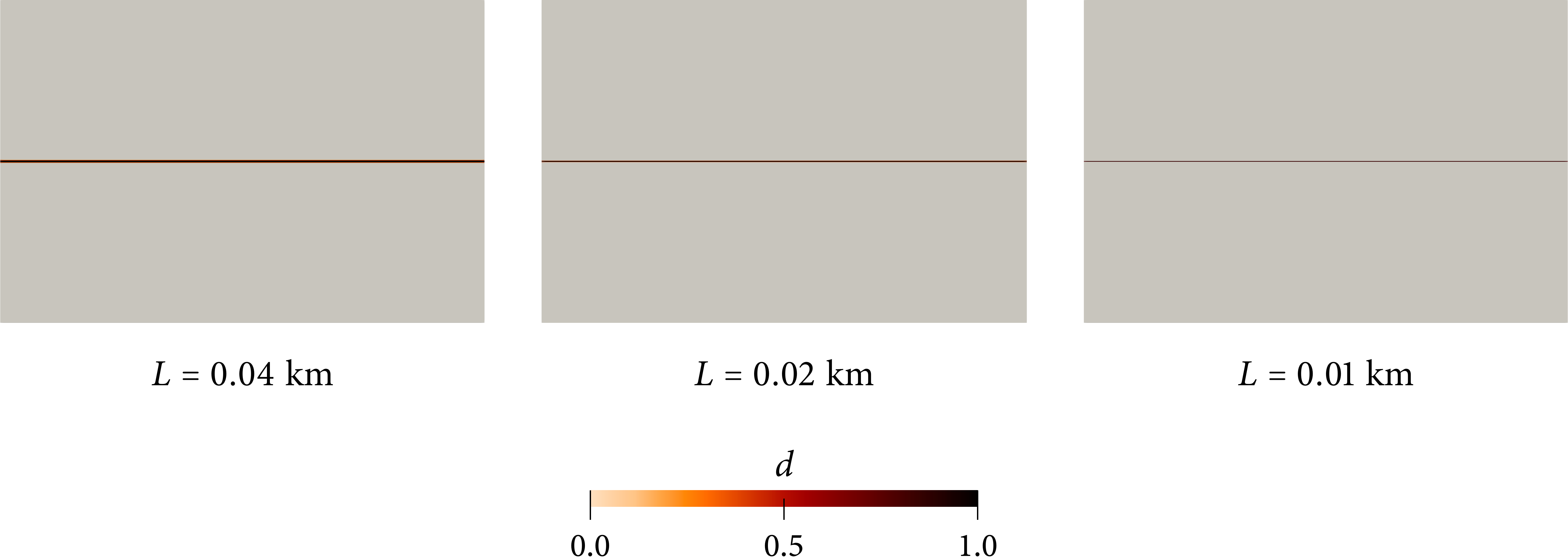}
  \caption{Rupture of a fault: phase-field approximations of the fault with three different length parameters.}
  \label{fig:fault-rupture-pf}
\end{figure}

To examine the dependence of the phase-field solutions on the length parameter, $L$, we repeatedly simulate the same example with three different length parameters, namely, $L = 0.04$ km, $0.02$ km, and $0.01$ km. 
Figure~\ref{fig:fault-rupture-pf} presents the phase-field approximations of the fault with these three length parameters.
Here, the initial phase-field distributions are constructed by prescribing high crack driving forces at the location of the fault, as in Borden~\etal~\cite{borden2012phase}. 
Also, to examine mesh sensitivity, we discretize the fault region ($d>0$) with three different levels of spatial discretization, namely, $L/h = 5$, $10$, and $20$, where $h$ denotes the element size.
The time step size is \revised{chosen according to the following criterion proposed by Lapusta~\etal~\cite{lapusta2000elastodynamic},
\begin{linenomath}
\begin{align}
    \Delta t = \min [\xi D_{c} /V_{i} ], 
\end{align}
\end{linenomath}
where $V_{i}$ is the slip rate at element $i$, and $\xi$ is a prescribed parameter. 
In essence, this equation ensures that the chosen time step size gives a slip increment that is smaller than a certain fraction $\xi$ of the characteristic slip distance $D_{c}$.
Here, we choose $\xi \approx 1/10$ to ensure sufficient accuracy and stability of numerical solutions.
Given the adopted characteristic slip distance $D_{c} = 0.02$ m, and the maximum slip rate $V_{\max} \approx 2$ m/s, we set $\Delta t = 0.001$ s.
}

Since the discontinuity is represented implicitly by the phase field, we assign the initial shear stress on the fault, $\tau_{f, \text{init}}=31.382$ MPa, through a preloading stage.
In this preloading stage, we apply an antiplane shear traction of the same magnitude as $\tau_{f, \text{init}}$ on the top boundary and solved the momentum balance equation once.
The displacement constraints on the top boundary are relaxed during this stage.
Once the shear stress is initialized, we constrain the top boundary again so that the initial shear stress is maintained on the fault. 

After initializing the shear stress, we trigger a rupture on the fault by reducing the reference friction coefficient $\mu_{0}$ inside the nucleation region as
\begin{linenomath}
\begin{align}
  \mu_{0} (x,t) = \mu_{0, \text{init}} - (\mu_{0, \text{init}} - \mu_{0, \min} )  G(x) H(t), 
\end{align}
\end{linenomath}
where $G(x)$ and $H(t)$ are spatial and temporal evolution functions, respectively, defined as 
\begin{linenomath}
\begin{align}
  G(x) =  \left \{
    \begin{array}{ll}
      \exp \left[
      \dfrac{(x - 30)^2}{(x - 30)^2 - 9} \right] & \text{if} \:\: 27\; \text{km} < x < 33 \; \text{km} ,  \\ 
      0 & \text{otherwise} , 
    \end{array}
  \right. 
\end{align}
\end{linenomath}
and 
\begin{linenomath}
\begin{align}
  H(t) = \left \{
    \begin{array}{ll}
     0 & \text{if} \:\: t = 0 \; \text{s} ,\\
      \exp\left[\dfrac{(t - 1)^2}{t(t - 2)} \right] & \text{if} \:\: 0 \; \text{s} < t < 1 \; \text{s} , \\ 
      1 & \text{if} \:\: t > 1 \; \text{s} . 
    \end{array}
  \right. 
\end{align}
\end{linenomath}

\begin{figure}[htbp]
    \centering
    \subfloat[]{\includegraphics[height=0.4\textwidth]{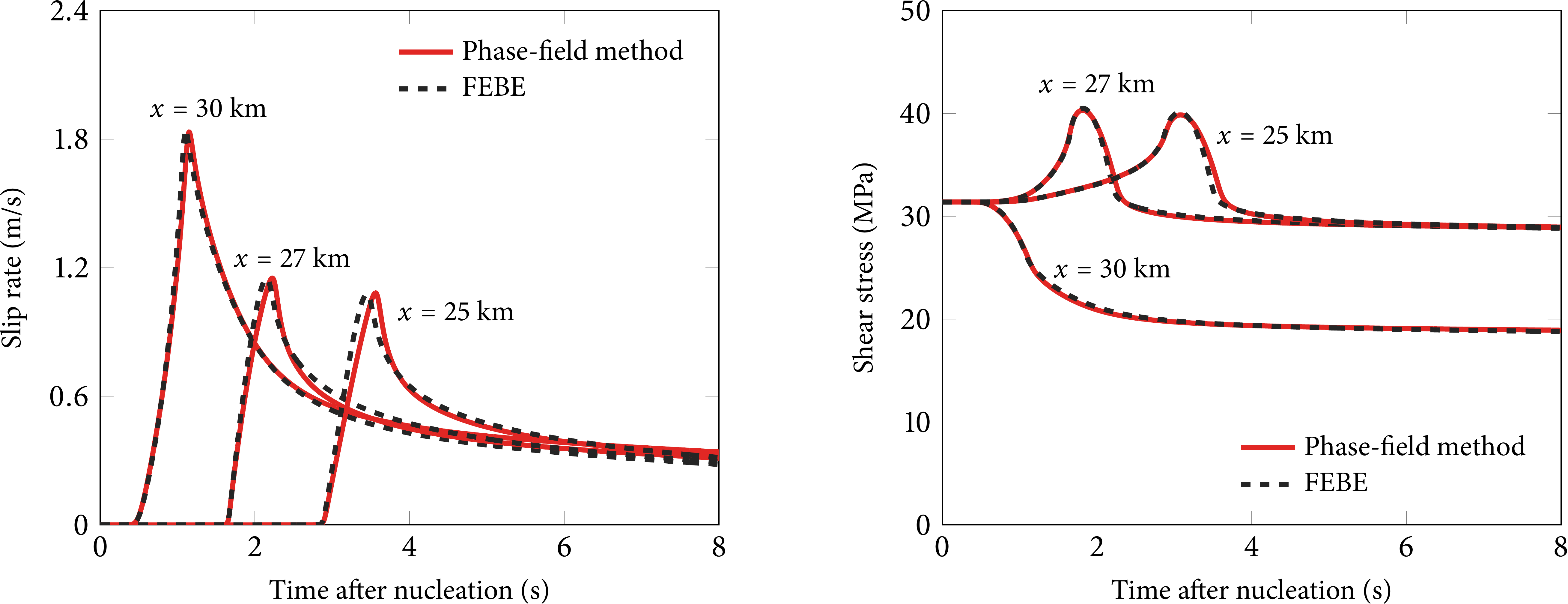} \label{fig:fault-rupture-compare}} \hspace{0.5em}
    \subfloat[]{\includegraphics[height=0.4\textwidth]{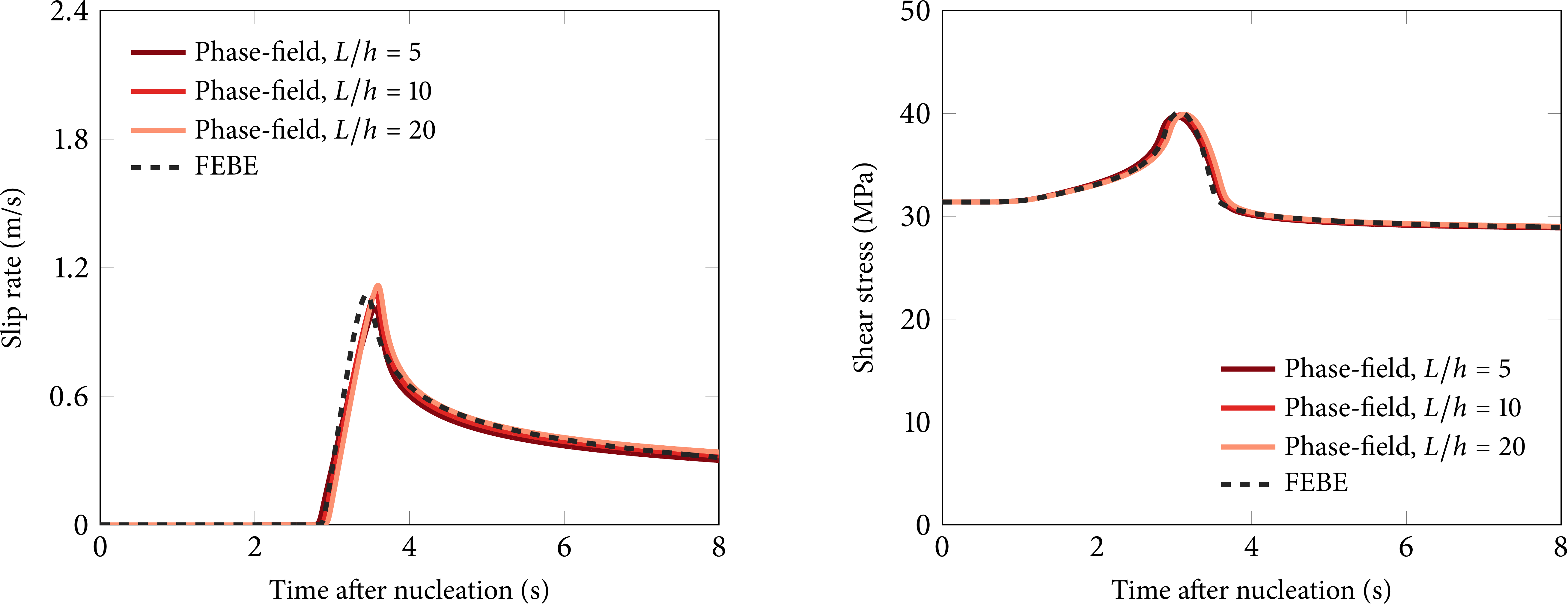} \label{fig:fault-rupture-mesh}} \hspace{0.5em}
    \subfloat[]{\includegraphics[height=0.4\textwidth]{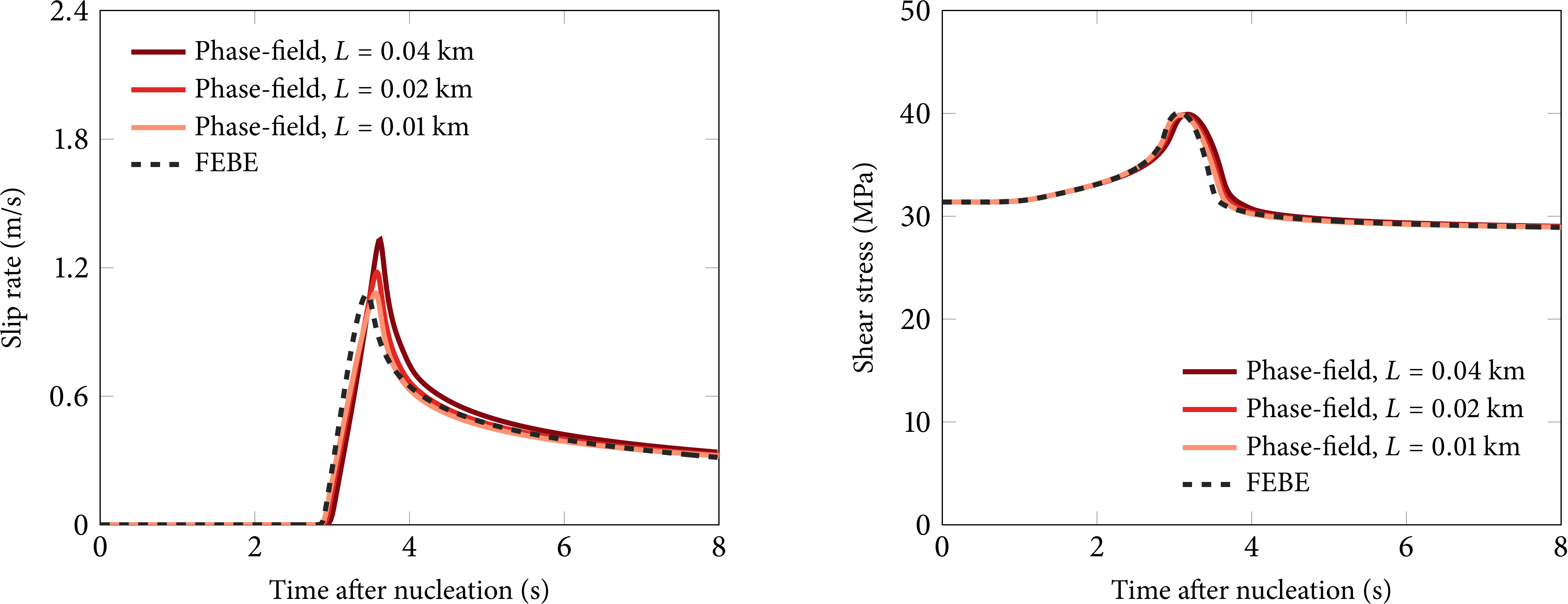} \label{fig:fault-rupture-length}} 
    \caption{Rupture of a fault:
    (a) comparison between the phase-field and FEBE results;
    (b) mesh convergence (the results at $x=25$ km are shown); 
    (c) length parameter convergence (the results at $x = 25$ km are shown). }
    \label{fig:fault-rupture}
\end{figure}

Figure~\ref{fig:fault-rupture-compare} presents the results of the phase-field and FEBE methods at three locations on the fault. (The phase-field solutions in this figure are obtained with $L = 0.01$ km and $L/h = 10$.) 
The phase-field results match well the results from the FEBE method in terms of the time evolution of the slip rate and shear stress. 
\revised{
At each location, the slip rate first increases and then decreases as rupture travels across the point.  
Also, after the nucleation of the rupture, the slip rates at the three locations reach their peak values in a sequential manner, indicating the propagation of the rupture along the fault. 
The shear stresses also evolve in a similar way, except at the nucleation location ($x = 30$ km) where the shear stress decreases at the beginning due to the reduction of friction.
All these results are consistent in the phase-field and FEBE solutions.}
Also, one can observe that the initial shear stresses at all three locations are correctly assigned as $\tau_{f, \text{init}}=31.382$ MPa. 
This good agreement verifies the proposed phase-field formulation and its implementation including stress initialization.

Next, in Fig.~\ref{fig:fault-rupture-mesh} we examine the mesh convergence of the phase-field method by comparing its solutions from different values of $h/L$ when $L = 0.01$ km. 
As shown, the numerical solutions converge to the FEBE solution as the element size becomes smaller.
This establishes the convergence of the phase-field formulation with mesh refinement, as long as the length parameter, $L$, is sufficiently discretized. 
The existence of such length parameter effectively regularizes the formulation and ensures mesh objectivity.
We emphasize that such mesh convergence is a distinct advantage of the phase-field method over the majority of existing continuous approaches to fault rupture.

In Fig.~\ref{fig:fault-rupture-length}, we further examine the sensitivity of the phase-field solutions to the phase-field length parameter, $L$, fixing $L/h = 10$,
One can see that as the length parameter decreases, the phase-field results become closer to the FEBE solution. 
It confirms that the phase-field approximation converges to the sharp discontinuity as $L$ decreases, which is a well-known property of phase-field modeling attained by the $\Gamma$-convergence of the crack density functional. 

\revised{
Before closing this verification example, we briefly recapitulate the advantages and disadvantages of the phase-field method compared with discontinuous methods such as FEBE. For problems where the fault geometry is predefined and fixed---like this verification example---the phase-field method may not be preferred to discontinuous methods, because the phase-field method entails much higher computational cost due to its use of very fine discretization along the discontinuity.
However, whenever the fault geometry may evolve (\eg~fault propagation, off-fault damage growth), the phase-field method has a significant advantage over discontinuous methods because the phase-field method can capture evolving fault geometry without sophisticated algorithms.
In what follows, we consider such problems for which the phase-field method can be highly beneficial.
}

\subsection{Propagation of a fault}

\begin{figure}[htbp]
  \centering
  \includegraphics[width=0.6\textwidth]{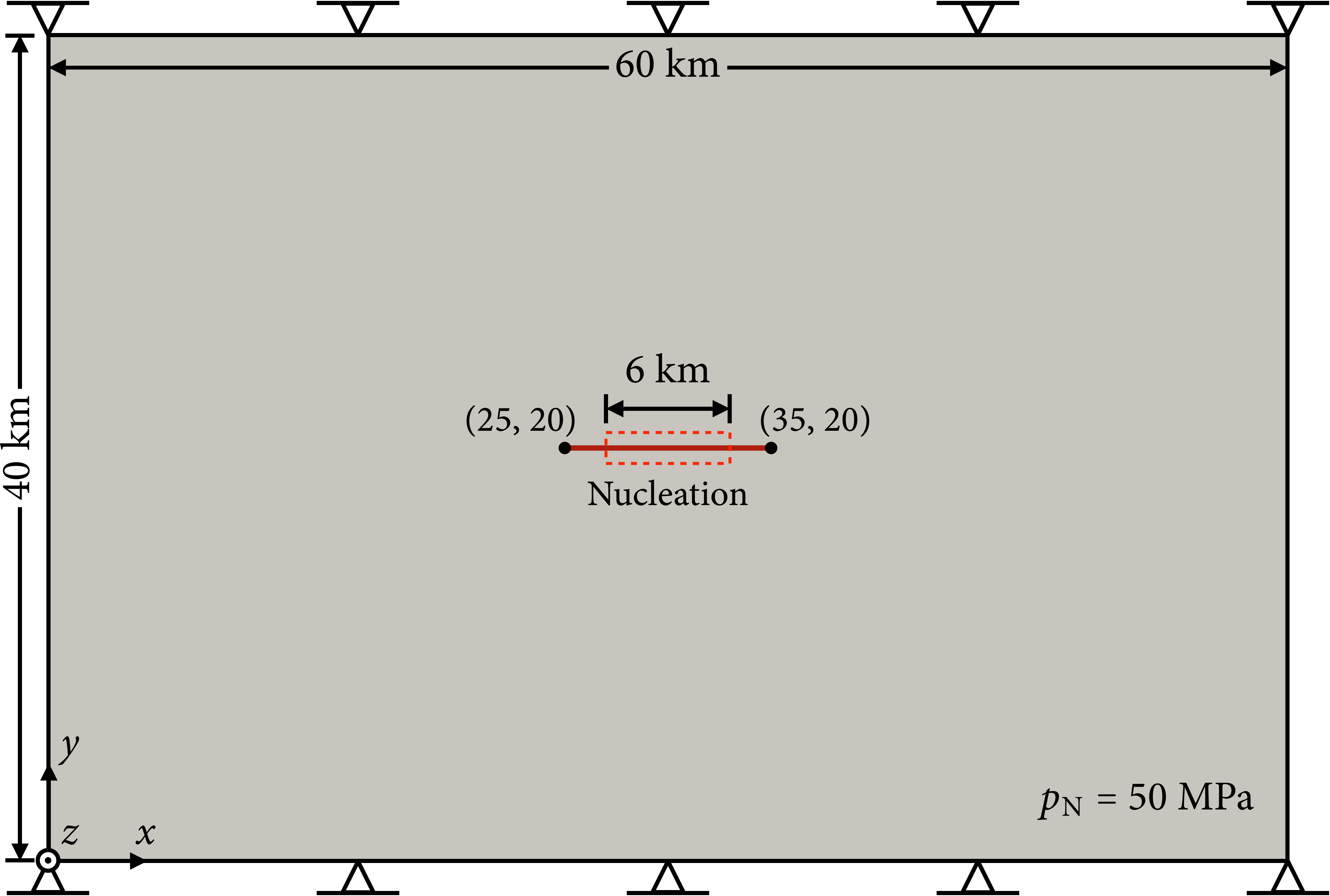}
  \caption{Propagation of a fault: geometry and boundary conditions.}
  \label{fig:single-fault-propagate-setup}
\end{figure}

Having verified the phase-field formulation, we now apply it to simulate quasi-dynamic propagation of a fault. 
For this purpose, we modify the previous example by shortening the fault such that it is allowed to propagate after the earthquake nucleation. 
The detailed setup is shown in Fig.~\ref{fig:single-fault-propagate-setup}. 
To model the fault propagation, we set the cohesion as $c = 3$ MPa and the critical fracture energy as $\mathcal{G}_{\rn{2}} = 1 \; \text{MJ/m}^2$.
The adopted critical fracture energy is within the range of the shear fracture energies inferred from real earthquakes, see Abercrombie and Rice~\cite{abercrombie2005can}. 
The length parameter is set to be $L = 0.04$ km, and the mesh is locally refined to satisfy $L/h = 10$ in the fault region.
Note that this combination of $L$ and $L/h$ has shown sufficient accuracy in the previous example.
The time step size is slightly increased to $\Delta{t} = 0.004$ s.  
We assign the initial stress and trigger earthquake nucleation in the same way as the previous example. 
The state variable at the newly developed slip surface is initialized according to the background slip rate $10^{-9}$ m/s with the steady state assumption. Alternative initialization procedures that account for possible deviation from steady state will be explored in future studies.  

\begin{figure}[htbp]
  \centering
  \includegraphics[width=0.8\textwidth]{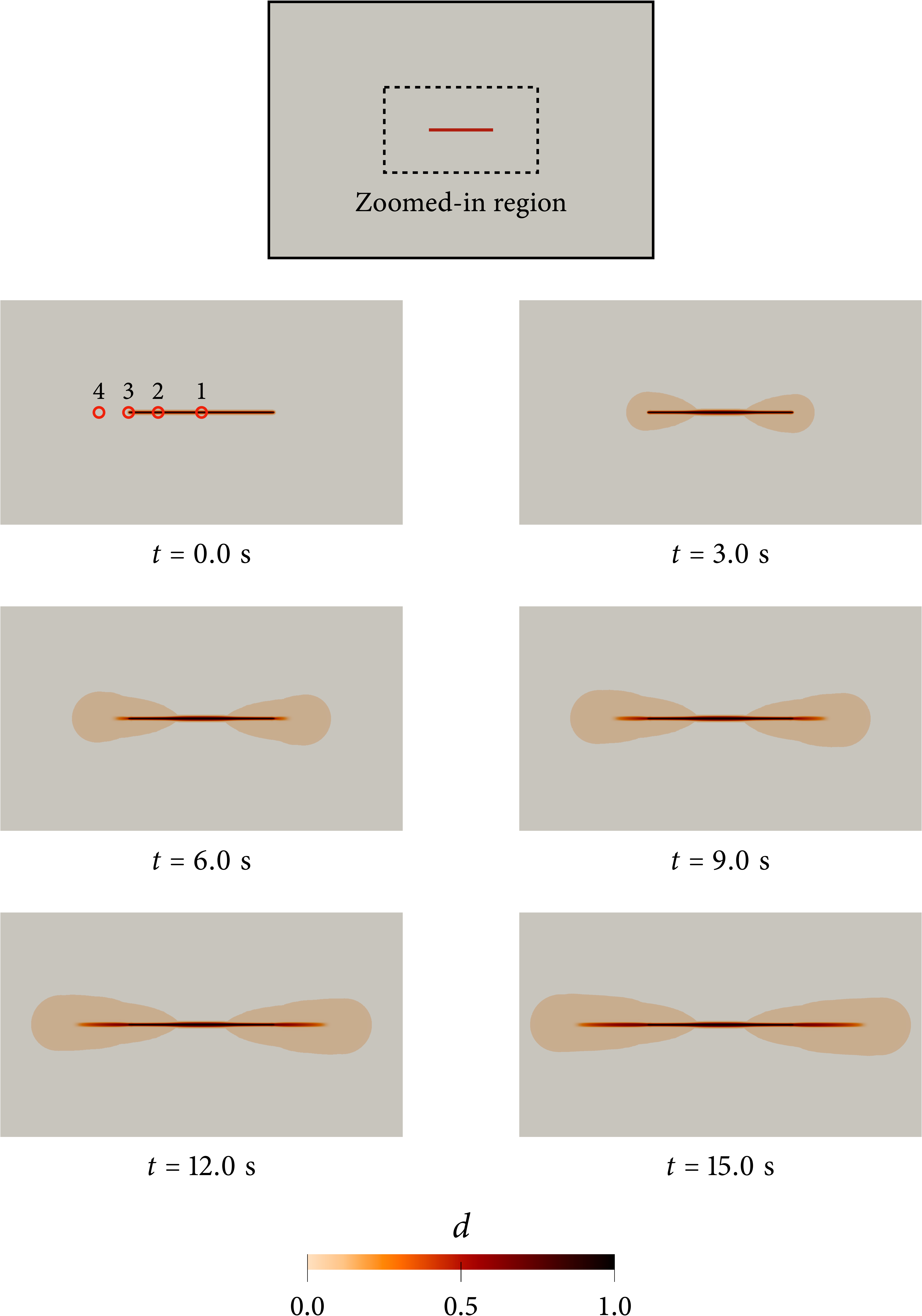}
  \caption{Propagation of a fault: evolution of the fault and off-fault damage. Point 1: $x = 30$ km. Point 2: $x = 27$ km. Point 3: $x = 25$ km. Point 4: $x = 23$ km.}
  \label{fig:single-fault-propagate-pf}
\end{figure}

Figure~\ref{fig:single-fault-propagate-pf} presents the phase-field simulations at different time steps after earthquake nucleation. 
One can see that in this two-dimensional antiplane setup, the fault propagates horizontally, being consistent with the analytical predictions provided by Poliakov~\etal~\cite{poliakov2002dynamic}.  
Remarkably, around the fault, regions with low phase-field values ($0<d<1$) emerge, which can be interpreted as the off-fault damage in rock masses due to fault rupture. 
The damage pattern is comparable to the one obtained in Hajarolasvadi and Elbanna~\cite{hajarolasvadi2017new}, whereby a plasticity model is used to simulate the off-fault damage for a two-dimensional antiplane fault. 
These results suggest that the proposed phase-field formulation can simulate not only fault propagation but also off-fault damage.

\begin{figure}[htbp]
    \centering
    \subfloat[Slip rate]{\includegraphics[height=0.4\textwidth]{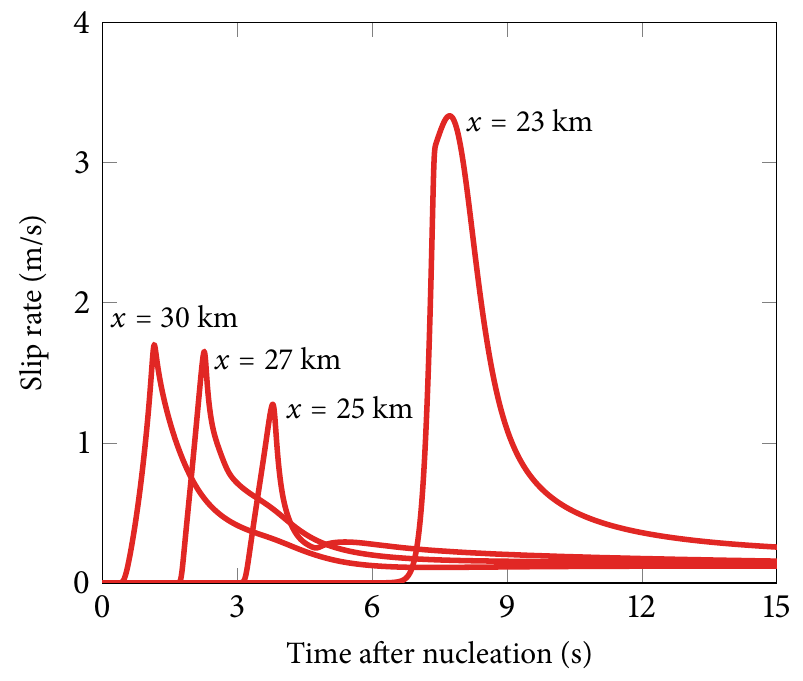}} \hspace{0.5em}
    \subfloat[Shear stress]{\includegraphics[height=0.4\textwidth]{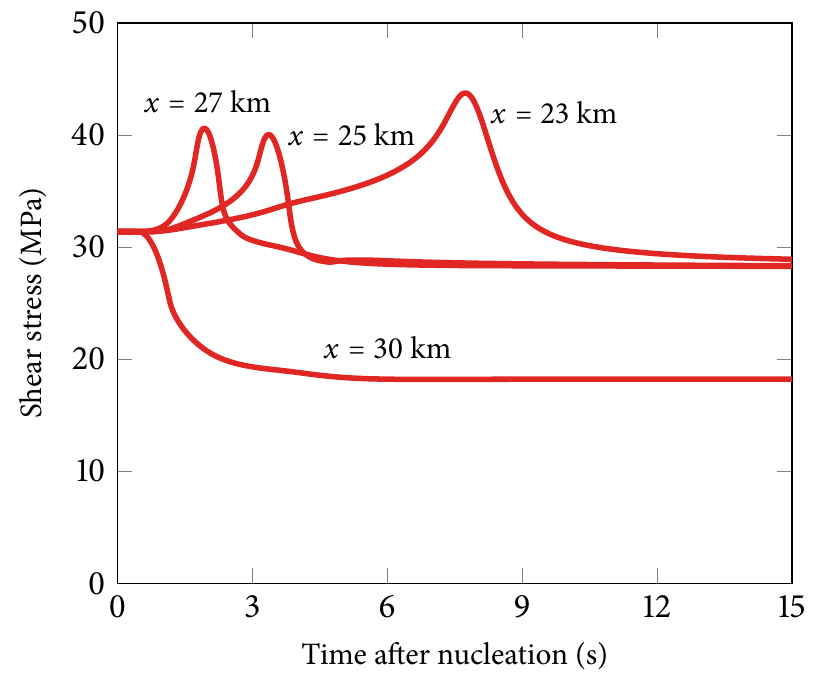}}
    \caption{Propagation of a fault: (a) slip rates and (b) shear stresses at four different locations on the propagating fault denoted in Fig.~\ref{fig:single-fault-propagate-pf}. }
    \label{fig:single-fault-propagate-slip-stress}
\end{figure}

To further investigate how the slip rate and shear stress evolve on the propagating fault, 
in Fig.~\ref{fig:single-fault-propagate-slip-stress}, we present the slip rates and shear stresses at four different locations along the plane passing through the fault, namely, the center of the fault ($x=30$ km), the boundary of the nucleation zone ($x=27$ km), the tip of the initial fault ($x=25$ km), and a location outside the initial fault ($x=23$ km). 
As shown, the slip rate at the fault center reaches its peak value first, and then those at other points reach their peaks as the rupture propagates outward. 
Also, it is noted that the peak slip rate at $x = 23$ km has the highest value among the peak slip rates at the four locations. 
This high peak slip rate would be the result of an increase in the crack driving force associated with the newly developed fault plane at $x = 23$ km. 

\subsection{Growth of faults and off-fault damage in a heterogeneous domain}

\begin{figure}[htbp]
    \centering
    \includegraphics[width=\textwidth]{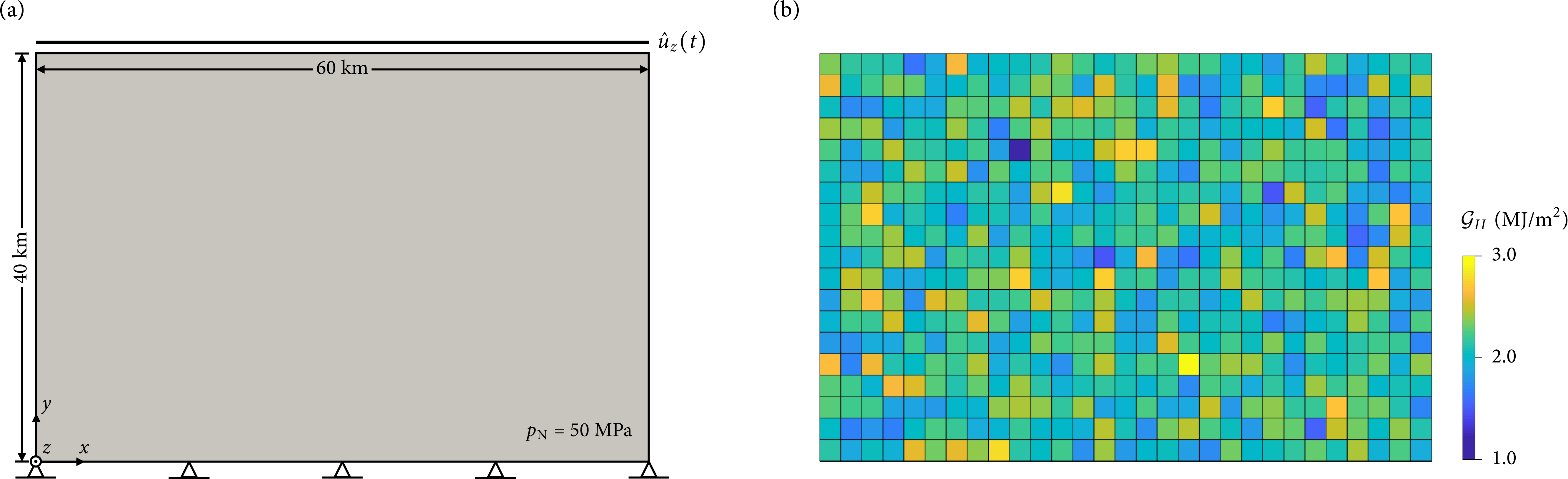}
    \caption{Growth of faults and off-fault damage in a heterogeneous domain: (a) geometry and boundary conditions, and (b) the distribution of shear fracture energy.}
    \label{fig:fault-heterogeneous-setup}
\end{figure}

In the last example, we apply the phase-field approach to simulate spontaneous growth of faults and off-fault damage in a heterogeneous domain, which is a highly challenging task for all the existing approaches to fault simulations. 
To this end, we again modify the previous example, removing the preexisting fault and assigning a random field of shear fracture energy within its realistic range~\cite{rice2005off}.
Figure~\ref{fig:fault-heterogeneous-setup} depicts the resulting problem configuration and the distribution of the shear fracture energy in the domain.
To induce shearing in the domain, we apply a prescribed antiplane displacement rate of $1.2$ m/s on the top boundary.
To simulate the problem with reasonable computational cost, we slightly increase the length parameter to $L = 0.1$ m.
The whole domain is then uniformly discretized to satisfy $L/h = 5$.
Note that local mesh refinement is no longer possible because the potential faulting regions are not known \textit{a priori}. 
Adaptive mesh refinement may be considered in future studies.
The time step size is set as $\Delta{t} = 0.01$ s.  
All the other aspects remain the same as the previous example.

\begin{figure}[htbp]
    \centering
    \subfloat[]{\includegraphics[width=0.85\textwidth]{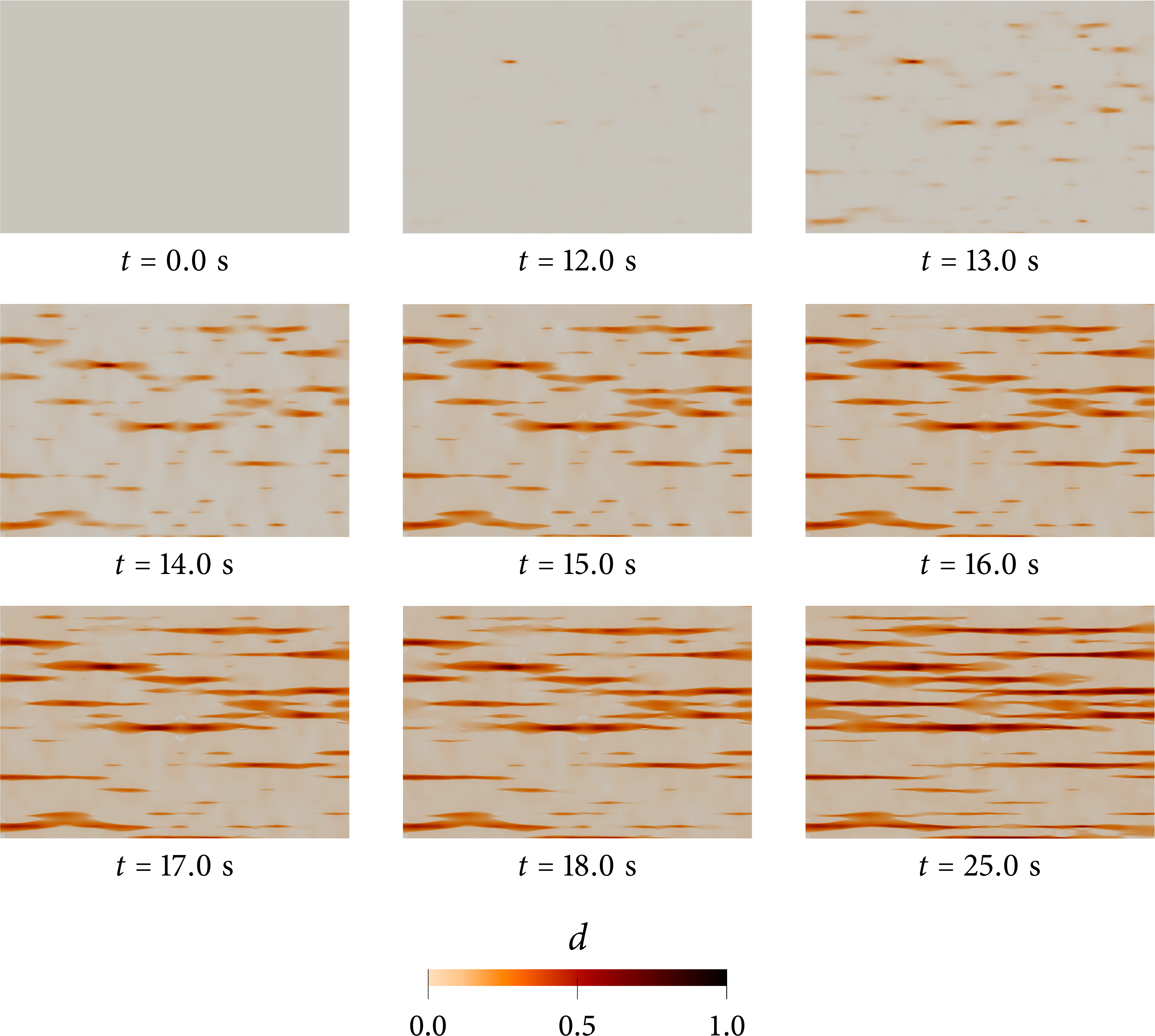} \label{fig:fault-heterogeneous-pf}}\\
    \subfloat[]{\includegraphics[width=0.5\textwidth]{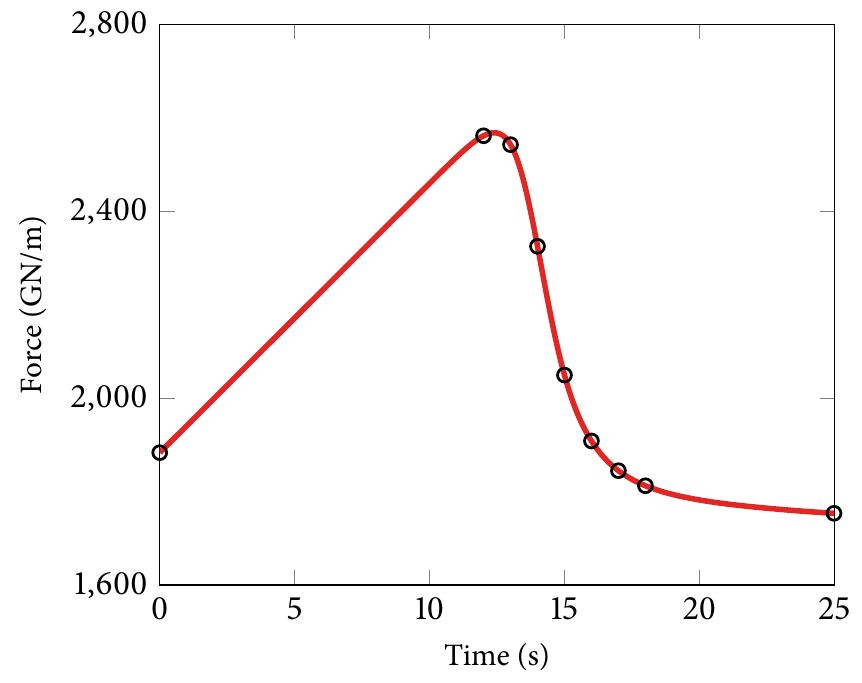} \label{fig:fault-heterogeneous-force}}
    \caption{Quasi-dynamic fault growth in a heterogeneous domain: (a) evolution of faults and off-fault damage; 
    (b) the evolution curve of the reaction force at the top boundary.}
\end{figure}

Figures~\ref{fig:fault-heterogeneous-pf} and~\ref{fig:fault-heterogeneous-force} present the time evolutions of the phase fields and the reaction forces, respectively. 
The time instances in Fig.~\ref{fig:fault-heterogeneous-pf} correspond to those marked in Fig.~\ref{fig:fault-heterogeneous-force}. 
Fault planes first emerge in the region where the shear fracture energy is low. 
As the domain becomes sheared further, the fault planes propagate progressively in the horizontal direction and coalesce with each other to form a larger plane.
These results suggest that the phase-field method naturally simulate the nucleation, propagation, and coalescence of faults in heterogeneous materials. 
Also, as shown in Fig.~\ref{fig:fault-heterogeneous-force}, the reaction force increases linearly in the beginning and then attains its peak value at $t \approx 12$ s, which is the moment when the first fault surface initiates as shown in Fig.~\ref{fig:fault-heterogeneous-pf}.  
Subsequently, the domain undergoes softening due to the progressive localization of slip surfaces and then reaches a residual state. 
This post-peak softening response is consistent with the strain-weakening behavior of fault gouge materials when slip surfaces are localized~\cite{marone1998laboratory,rathbun2010effect}, which to some extent validates the proposed phase-field method.

\begin{figure}[htbp]
    \centering
    \includegraphics[width=0.9\textwidth]{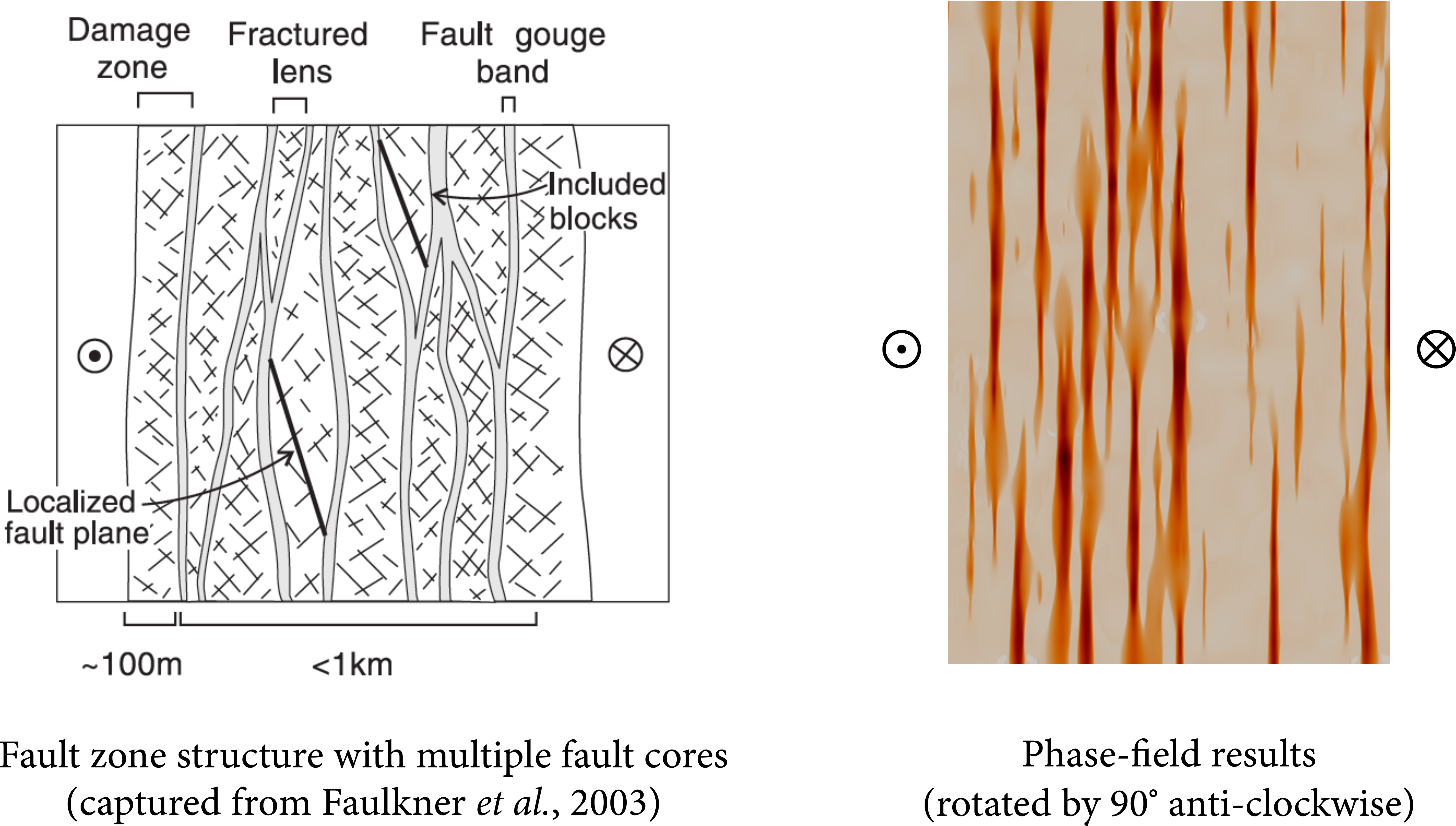}
    \caption{Growth of faults and off-fault damage in a heterogeneous domain: comparison between a conceptual fault zone model and the phase-field simulation results.}
    \label{fig:fault-heterogeneous-comparison}
\end{figure}

Furthermore, in Fig.~\ref{fig:fault-heterogeneous-comparison} we compare the phase-field simulation results with a conceptual model of typical strike-slip fault zone structure in Faulkner~\etal~\cite{faulkner2003internal}.
Despite the difference in the domain scales, the simulated patterns of faults and damage are analogous to the multiple strands of fault slip surfaces in the conceptual model. 
It is also noted that the development of multiple slip surfaces in the simulations may be attributed to that the slip surfaces are triggered by material heterogeneity alone without any preexisting faults.
In other words, because slip surfaces are no longer restricted to any predefined fault planes, they can nucleate freely in regions of lower fracture energies.
This observation may provide insights into the mechanical origin of fault zones exhibiting such multiple slip surfaces, \eg~Carboneras fault in southeastern Spain~\cite{faulkner2003internal,faulkner2008structure}.

\section{Conclusion}
\label{sec:summary}

A phase-field model for quasi-dynamic fault rupture and propagation has been proposed, in which the fault surface is diffusely approximated by the phase-field variable.
Built on a recently developed phase-field method for quasi-static shear fracture with frictional contact, the proposed formulation has incorporated rate- and state-dependent friction, radiation damping, and their impacts on the fault rupture and propagation process.
We have verified that the phase-field solutions converge to the results of a discontinuous approach, and are consistent with rate-and-state friction laws, unlike several of the existing continuous approximation approaches to numerical modeling of fault ruptures. 
In addition, we have demonstrated that the phase-field approach can simulate complex growth of faults and off-fault damage, without any sophisticated algorithms for tracking geometry nor an additional consideration for damage evolution.
The phase-field approach may thus provide new opportunities for numerical investigations of complex earthquakes that have been overly challenging for the existing numerical approaches. 
Future work will focus on relaxing the assumptions made in this initial work, namely, the antiplane kinematics and the quasi-dynamic approximation.
\revised{To this end, we will fully incorporate the tensorial nature of stress and strain and inertial terms into the formulation.}

\section*{Acknowledgments}
This work was supported by the National Research Foundation of Korea (NRF) grant funded by the Korean government (MSIT) (No. 2022R1F1A1065418).
A.E.E. acknowledges support from the Southern California Earthquake Center through a collaborative agreement between NSF Grant No. EAR0529922 and USGS Grant No. 07HQAG0008, and the NSF CAREER Award No. 1753249 for modeling complex fault zone structures. 
M.S.M and A.E.E acknowledge additional support from the U.S. Department of Energy under Award No. DE-FE0031685 for modeling damage in fault zones during natural and induced seismicity.
Portions of this work were performed under the auspices of the U.S. Department of Energy by Lawrence Livermore National Laboratory under Contract DE-AC52-07NA27344.

\appendix

\section{Finite element discretization and solution algorithms}
\label{appendix:discretization}

This appendix describes how to apply a standard Galerkin finite element method for solving the phase-field formulation. 

\subsection{Variational formulation}
To develop a variational formulation of the governing equations~\eqref{eq:momentum-balance-quasidynamic-antiplane} and~\eqref{eq:pf-evolution-quasidynamic-antiplane}, we first define the trial solution spaces for the antiplane displacement field $u_{z}$ and the phase field $d$ as 
\begin{linenomath}
\begin{align}
	\mathcal{S}_{u} &:= \left \{ u_{z} \rvert u_{z} \in H^{1} , \; u_{z} = \hat{u}_{z} \, \, \text{on} \; \pd_{u} \Omega  \right \} ,  \\
	\mathcal{S}_{d} &:= \left \{d  \rvert d \in H^{1} \right\}   , 
\end{align}
\end{linenomath}
where $H^{1}$ denotes the Sobolev space of order one. 
The spaces of variations for the antiplane displacement field $u_{z}$ and the phase field $d$ are defined as 
\begin{linenomath}
\begin{align}
	\mathcal{V}_{u} &:= \left \{ \eta \rvert \eta \in H^{1} , \; \eta = 0 \, \, \text{on} \; \pd_{u} \Omega  \right \} ,  \\
	\mathcal{V}_{d} &:= \left \{\phi  \rvert \phi \in H^{1} \right\}   . 
\end{align}
\end{linenomath}
Following the standard procedure, we obtain the discrete variational form of Eqs.~\eqref{eq:momentum-balance-quasidynamic-antiplane} and~\eqref{eq:pf-evolution-quasidynamic-antiplane} as
\begin{linenomath}
\begin{align}
	\mathcal{R}^{h}_{u} & := - \int_{\Omega} \grad \eta^{h} \cdot \tensor{\tau} \: \dd V + \int_{\pd_{t} \Omega} \eta^{h} \hat{t}_{z} \: \dd A = 0 , \label{eq:variational-momentum-balance} \\
	\mathcal{R}^{h}_{d} & := \int_{\Omega} \phi^{h} g'(d) \mathcal{H}^{+} \: \dd V + \int_{\Omega} \dfrac{3 \mathcal{G}_{\rn{2}}}{8L} (2L^2 \grad \phi^{h} \cdot \grad d^{h} + \phi^{h}) \: \dd V = 0, \label{eq:variational-pf-evolution}
\end{align}
\end{linenomath}
where the superscript $h$ denotes spatially discretized quantities. 
In the above, the variational equations are defined as residuals, $\mathcal{R}^{h}_{u}$ and $\mathcal{R}^{h}_{d}$, so that each of them can be solved by Newton's method.

\subsection{Staggered solution scheme}
To solve Eqs.~\eqref{eq:variational-momentum-balance} and \eqref{eq:variational-pf-evolution}, we make use of a staggered algorithm that has been commonly adopted in the phase-field literature~\cite{miehe2010phase,geelen2019phase,fei2020phaseb,fei2021double}. 
Specifically, in each time step, we first solve Eq.~\eqref{eq:variational-momentum-balance} for the antiplane displacement $u_{z}$ fixing the phase-field variable $d$, then solve Eq.~\eqref{eq:variational-pf-evolution} for the phase-field variable $d$ fixing the displacement field $u_{z}$.  
The main advantage of this staggered algorithm over a monolithic solution algorithm is that it significantly improves the numerical robustness without much compromise in accuracy as long as the time step is sufficiently small.

Since the discrete variational equations~\eqref{eq:variational-momentum-balance} and \eqref{eq:variational-pf-evolution} are incrementally nonlinear, we use Newton's method to solve each equation within a staggered iteration step. 
To solve the antiplane displacement $u_{z}$, we linearize Eq.~\eqref{eq:variational-momentum-balance} as
\begin{linenomath}
\begin{align}
	\delta \mathcal{R}^{h}_{u} := \int_{\Omega} \grad \eta^{h} \cdot \mathbb{C} \cdot \grad \delta u^{h}_{z} \: \dd V , \label{eq:linearize-momentum-balance}
\end{align}
\end{linenomath}
where $\delta$ denotes the linearization operator, and $\mathbb{C}$ is the stress-strain tangent operator.
We solve this linearized equation for the displacement increment $\Delta u^{h}_{z}$, with a fixed phase-field at each Newton iteration step. 
After calculating the displacement field, we use Newton's method to solve the linearized version of Eq.~\eqref{eq:variational-pf-evolution}, given by
\begin{linenomath}
\begin{align}
	\delta \mathcal{R}^{h}_{d} := \int_{\Omega} \phi^{h} g''(d) \mathcal{H}^{+} \delta d^{h} \: \dd V + \int_{\Omega} \dfrac{3\mathcal{G}_{\rn{2}}}{4} L \grad \phi^{h} \cdot \grad \delta d^{h} \: \dd V .  \label{eq:linearize-pf-evolution}
\end{align}
\end{linenomath}
It is noted that the crack driving force $\mathcal{H}^{+}$ is updated in the displacement solution stage, while it remains constant in the phase-field solution stage.  

\subsection{Material update algorithm}
To evaluate the discretized variational equations~\eqref{eq:variational-momentum-balance} and \eqref{eq:variational-pf-evolution} and their linearization~\eqref{eq:linearize-momentum-balance} and \eqref{eq:linearize-pf-evolution}, we have to compute the internal variables -- the crack driving force, shear stress tensor, and the consistent tangent operator -- at each material point.
Algorithm~\ref{algo:material-update} presents our procedure to update these internal variables from the previous time step $t_{n}$ to the next time step $t_{n+1}$.
(For simplicity, quantities at $t_{n+1}$ are denoted without any additional subscript, while quantities at $t_{n}$ are denoted with subscript $n$). 
The algorithm calculates the crack driving force, the stress tensor, and the consistent tangent operator according to the damage and contact condition detected on the material point of interest.

Three important points in the material update algorithm deserve elaboration.
First, when the slip rate is zero under an intact ($d = 0$) or a stick condition, we update the friction coefficient with the original Dieterich--Ruina law~\eqref{eq:rs-friction-original}, which gives $\mu = \mu_{0}$. 
This is because the regularized Dieterich--Ruina law~\eqref{eq:rs-friction-regularize} gives an unrealistic zero friction coefficient when the slip rate is zero. 
Second, we apply the implicit Euler method to update the state variable.
So the discrete form of the aging law~\eqref{eq:aging-law} is written as
\begin{linenomath}
\begin{align}
  \dot{\theta} = \dfrac{\theta - \theta_{n}}{\Delta t} = 1 - \dfrac{V \theta}{D_{c}} \quad \text{with} \; \Delta t := t_{n+1} - t_{n}. \label{eq:aging-law-implicit-euler}
\end{align}
\end{linenomath} 
Here, the slip rate is approximated by 
\begin{linenomath}
\begin{align}
  V := \dot{\zeta}_{f} = \dfrac{\zeta_{f} - (\zeta_{f})_{n}}{\Delta t } = \dfrac{\Delta \zeta_{f}}{\Delta t} . \label{eq:slip-rate-discrete}
\end{align}
\end{linenomath} 
Rearranging Eq.~\eqref{eq:aging-law-implicit-euler} thus gives 
\begin{linenomath}
\begin{align}
  \theta = \dfrac{\theta_{n} + \Delta t}{1 + \Delta t V/D_{c}} . \label{eq:state-discrete}
\end{align}
\end{linenomath} 
Third, we fix the degradation function $g(d)$ and the threshold crack driving force $\mathcal{H}_{t}$ throughout the problem.
This is because $g(d)$ and $\mathcal{H}_{t}$ are constant in the present formulation due to the assumption that the peak and residual friction coefficients are identical, see Eqs.~\eqref{eq:gd} and~\eqref{eq:H-threshold}.
If the two friction coefficients are different, we can update the degradation function and the crack driving force as in Fei and Choo~\cite{fei2020phaseb}.

\begin{algorithm}[htbp!]
    \setstretch{1.3}
    \caption{Material point update procedure for the proposed phase-field formulation.}
      \begin{algorithmic}[1]
        \Require $\Delta \tensor{\gamma}$, $d$, and $\tensor{n}$ at $t_{n+1}$.
        \State Calculate $\tensor{\gamma} = \tensor{\gamma}_{n} + \Delta \tensor{\gamma}$, $\Delta \gamma =  \Delta \tensor{\gamma} \cdot \tensor{n}$, $\tensor{\tau}_{m} = G \tensor{\gamma}$, and $\tau_{m} =  \tensor{\tau}_{m} \cdot \tensor{n} $
        \If {$d = 0$}
          \State Intact material. 
          \State Update $\zeta_{f} = 0$, $\tensor{\tau} = \tensor{\tau}_{m}$ and $\mathbb{C} = G \tensor{1}$. 
          \State Update $\tau_{p} = c + p_{\cn} \mu$ and $\tau_{r} = p_{\cn} \mu$, where $\mu = \mu_{0}$. 
          \If {$\lvert \tau_{m} \rvert > \tau_{p}$}
            \State Update $\mathcal{H}^{+} = \mathcal{H}_{t} + (\lvert \tau_{m} \rvert - \tau_{r}) \lvert \Delta \gamma \rvert$. 
          \Else
            \State Set $\mathcal{H}^{+} = \mathcal{H}_{t}$. 
          \EndIf
        \Else
          \State Update the trial stress in the fractured material $\tensor{\tau}^{\mathrm{tr}}_{f} = (\tensor{\tau}_{f})_{n} + G \Delta \tensor{\gamma}$.
          \State Update $\theta = (\theta_{n} + \Delta t)/(1 + \Delta t V_{n}/D_{c})$, and $\mu = \mu (V_{n}, \theta)$. 
          \State Calculate the yield function $F = \lvert \tau_{f}^\mathrm{tr} \rvert - p_{\cn} \mu - \eta V_{n}$, where $\tau_{f}^\mathrm{tr} = \tensor{\tau}_{f}^\mathrm{tr} \cdot \tensor{n}$. 
          \If {$F < 0$}
            \State Stick state. 
            \State Update $V = 0$, $\theta = \theta_{n} + \Delta t$, $\mu = \mu_{0}$, $\zeta_{f} = (\zeta_{f})_{n}$, $\tensor{\tau}_{f} = \tensor{\tau}_{f}^\mathrm{tr}$, and $\mathbb{C}_{f} = G \tensor{1}$. 
            \State Set $\mathcal{H}^{+} = \max_{t \in [0, t_{\max}]}\mathcal{H}^{+}(t)$. 
          \Else 
            \State Slip state. 
            \State Update $\alpha = \tau^\mathrm{tr}_{f} / \lvert \tau^\mathrm{tr}_{f} \rvert $. 
            \State Perform return mapping to update $V$, $\theta$, $\zeta_{f}$, $\mu$, $\tensor{\tau}_{f}$, and $\mathbb{C}_{f}$. 
            \State Update $\mathcal{H}^{+} = (\mathcal{H}^{+})_{n} + (\lvert \tau_{m} \rvert - \tau_{r} - \eta V)\lvert \Delta \gamma \rvert $, where $\tau_{r} = p_{\cn} \mu$. 
          \EndIf
          \State Update $\tensor{\tau} = g(d) \tensor{\tau}_{m} + [1 - g(d)] \tensor{\tau}_{f}$, and $\mathbb{C} = g(d) G \tensor{1} + [1 - g(d)] \mathbb{C}_{f}$.
        \EndIf
        \Ensure $\tensor{\tau}$, $\mathbb{C}$, and $\mathcal{H}^{+}$ at $t_{n+1}$.
      \end{algorithmic}
  \label{algo:material-update}
\end{algorithm}

\subsection{Return mapping and consistent tangent operator}
As shown in Algorithm~\ref{algo:material-update}, when the material point is in a slip state, the stress and the tangent operator in the fractured material, $\tensor{\tau}_{f}$ and $\mathbb{C}_{f}$, are updated through a return mapping procedure. 
The return mapping procedure is detailed below.

To begin, we define a vector of unknowns, namely, the shear stress tensor in the fractured material and the increment of the slip magnitude, as
\begin{linenomath}
\begin{align}
  \tensor{x} = 
  \begin{bmatrix}
    (\tensor{\tau}_{f})_{2 \times 1} \\ 
    \Delta \zeta_{f} 
  \end{bmatrix}_{3 \times 1} . 
\end{align}
\end{linenomath}
To solve the unknowns using Newton's method, we write the equations to be satisfied as a residual vector 
\begin{linenomath}
\begin{align}
  \tensor{r} = 
  \begin{bmatrix}
    (\tensor{\tau}_{f} + G\alpha \Delta \zeta_{f} \tensor{n} \Gamma_{d}(d, \grad d) - \tensor{\tau}_{f}^\mathrm{tr})_{2 \times 1} \\ 
    F 
  \end{bmatrix}_{3 \times 1}
   \rightarrow \tensor{0} . \label{eq:return-mapping-residual}
\end{align}
\end{linenomath}
At each Newton iteration, we solve for the increment of the unknowns vector, $\Delta \tensor{x}$, as
\begin{linenomath}
\begin{align}
  -\tensor{J} \cdot \Delta \tensor{x} = \tensor{r} \rightarrow 0 , 
\end{align}
\end{linenomath}
where $\tensor{J}$ is the Jacobian matrix, given by 
\begin{linenomath}
\begin{align}
  \tensor{J}  = 
  \begin{bmatrix}
    \tensor{1}_{2 \times 2} & \left(G\alpha \tensor{n} \Gamma_{d}(d, \grad d) \right)_{2 \times 1} \\ 
    \left(\dfrac{\tau_{f}}{\lvert \tensor{\tau}_{f} \cdot \tensor{n} \rvert } \tensor{n} \right)^{\intercal}_{1 \times 2} & - \left(p_{\cn}\dfrac{\dd \mu }{\dd V} + \eta \right) \dfrac{\pd V}{\pd \Delta \zeta_{f}}
  \end{bmatrix}_{3 \times 3} . 
\end{align}
\end{linenomath}
Here, $\tensor{1}$ denotes the second-order identity tensor. 
The derivative of the slip rate with respect to the increment of the slip magnitude can be calculated by taking the derivatives on both sides of Eq.~\eqref{eq:slip-rate-discrete}, which gives 
\begin{linenomath}
\begin{align}
  \dfrac{\pd V}{\pd \Delta \zeta_{f}} = \dfrac{1}{\Delta t} . 
\end{align}
\end{linenomath}
To complete the Jacobian matrix, we still need to determine the derivative of the friction coefficient with respect to the slip rate $\dd \mu / \dd V$.
We first extend this derivative using the chain rule
\begin{linenomath}
\begin{align}
  \dfrac{\dd \mu}{\dd V} = \dfrac{\pd \mu}{\pd V} + \dfrac{\pd \mu}{\pd \theta} \dfrac{\pd \theta}{\pd V}. \label{eq:d-mu-d-v}
\end{align}
\end{linenomath}
According to the regularized rate- and state-dependent friction law~\eqref{eq:rs-friction-regularize}, we obtain 
\begin{linenomath}
\begin{align}
  \dfrac{\pd \mu}{\pd V} = \dfrac{aK}{\sqrt{1 + (KV)^{2}}} , \label{eq:pd-mu-pd-v}
\end{align}
\end{linenomath}
and 
\begin{linenomath}
\begin{align}
  \dfrac{\pd \mu}{\pd \theta} = \dfrac{bKV}{\theta\sqrt{1 + (KV)^{2}}}, \label{eq:pd-mu-pd-theta}
\end{align}
\end{linenomath}
where 
\begin{linenomath}
\begin{align}
  K = \dfrac{1}{2V_{0}} \exp \left [\dfrac{\mu_{0} + b \ln(V_{0}\theta/D_{c})}{a} \right] . 
\end{align}
\end{linenomath}
We then take the derivative with respect to the slip rate on both sides of Eq.~\eqref{eq:state-discrete} and obtain 
\begin{linenomath}
\begin{align}
  \dfrac{\pd \theta}{\pd V} = - \dfrac{D_{c}\Delta t (\theta_{n} + \Delta t)}{(D_{c} + \Delta t V)^2} .  \label{eq:pd-theta-pd-v}
\end{align}
\end{linenomath}
Inserting Eqs.~\eqref{eq:pd-mu-pd-v}, \eqref{eq:pd-mu-pd-theta}, and \eqref{eq:pd-theta-pd-v} into Eq.~\eqref{eq:d-mu-d-v} gives 
\begin{linenomath}
\begin{align}
  \dfrac{\dd \mu}{\dd V} = \dfrac{aK}{\sqrt{1 + (KV)^{2}}} - \dfrac{bKV}{\theta\sqrt{1 + (KV)^{2}}}\dfrac{D_{c}\Delta t (\theta_{n} + \Delta t)}{(D_{c} + \Delta t V)^2} . \label{eq:d-mu-d-v-final}
\end{align}
\end{linenomath}

The consistent tangent operator in the fractured material is defined as
\begin{linenomath}
\begin{align}
  \mathbb{C}_{f} := \dfrac{\pd \tensor{\tau}_{f}}{\pd \tensor{\gamma}} . 
\end{align}
\end{linenomath}
To derive a closed-form expression for $\mathbb{C}_{f}$, we first linearize Eq.~\eqref{eq:return-mapping-residual} assuming that the residual reaches zero, and obtain
\begin{linenomath}
\begin{align}
   \begin{bmatrix}
    \left ( \delta \tensor{\tau}_{f} + G \alpha \tensor{n} \Gamma_{d}(d , \grad d) \delta \Delta \zeta_{f} \right)_{2 \times 1} \\ 
    \left(\dfrac{\tau_{f}}{\lvert \tensor{\tau}_{f} \cdot \tensor{n} \rvert } \tensor{n} \right) \cdot \delta \tensor{\tau}_{f} - \left(p_{\cn}\dfrac{\dd \mu }{\dd V} + \eta \right) \dfrac{1}{\Delta t} \delta \Delta \zeta_{f}
   \end{bmatrix}_{3 \times 1}
   = 
   \begin{bmatrix}
    (\delta \tensor{\tau}^\mathrm{tr}_{f})_{2 \times 1} \\ 
    0
   \end{bmatrix}_{3 \times 1} . \label{eq:residual-linearize}
\end{align}
\end{linenomath} 
We then linearize $\tensor{\tau}^\mathrm{tr}_{f} = (\tensor{\tau}_{f})_{n} + G (\tensor{\gamma} - \tensor{\gamma}_{n})$ considering $(\tensor{\tau}_{f})_{n}$ and $\tensor{\gamma}_{n}$ are constant during the update procedure. 
This operation gives 
\begin{linenomath}
\begin{align}
  \delta \tensor{\tau}^\mathrm{tr}_{f} = G \delta \tensor{\gamma} . 
\end{align}
\end{linenomath}
We then insert the above equation into Eq.~\eqref{eq:residual-linearize} and get 
\begin{linenomath}
\begin{align}
  \delta \tensor{\tau}_{f} + G\alpha \tensor{n} \Gamma_{d}(d , \grad d) \delta \Delta \zeta_{f} &= G\delta \tensor{\gamma} , \label{eq:strain-balance-linearize} \\ 
  \left(\dfrac{\tau_{f}}{\lvert \tensor{\tau}_{f} \cdot \tensor{n} \rvert } \tensor{n} \right) \cdot \delta \tensor{\tau}_{f} - \left(p_{\cn}\dfrac{\dd \mu }{\dd V} + \eta \right) \dfrac{1}{\Delta t} \delta \Delta \zeta_{f} &= 0 . \label{eq:yield-function-linearize}
\end{align}
\end{linenomath}
Rearranging Eq.~\eqref{eq:strain-balance-linearize} gives 
\begin{linenomath}
\begin{align}
  \delta \tensor{\tau}_{f} = G [\delta \tensor{\gamma} - \alpha \tensor{n} \Gamma_{d}(d , \grad d) \delta \Delta \zeta_{f}] \label{eq:linearized-tau-f}
\end{align}
\end{linenomath}
We then differentiate both sides of Eq.~\eqref{eq:linearized-tau-f} with respect to the total strain $\tensor{\gamma}$ and obtain
\begin{linenomath}
\begin{align}
  \mathbb{C}_{f} := \dfrac{\pd \tensor{\tau}_{f}}{\pd \tensor{\gamma}} = G \left[\tensor{1} - \alpha  \Gamma_d(d, \grad d) \left(\tensor{n} \dyadic\dfrac{\pd \Delta \zeta_{f}}{\pd \tensor{\gamma}}\right) \right] .  \label{eq:tangent-operator-unfinished}
\end{align}
\end{linenomath}
The remaining task is to determine the derivative of the slip magnitude increment with respect to the total strain. 
For this purpose, we insert Eq.~\eqref{eq:linearized-tau-f} into Eq.~\eqref{eq:yield-function-linearize}, and get 
\begin{linenomath}
\begin{align}
  G \left(\dfrac{\tau_{f}}{\lvert \tensor{\tau}_{f} \cdot \tensor{n} \rvert } \tensor{n} \right)\cdot \delta \tensor{\gamma} = \left[ G \alpha \dfrac{\tau_{f}}{\lvert \tensor{\tau}_{f} \cdot \tensor{n} \rvert } \Gamma_d(d, \grad d) + \left(p_{\cn}\dfrac{\dd \mu }{\dd V} + \eta \right) \dfrac{1}{\Delta t} \right] \delta \zeta_{f} . 
\end{align}
\end{linenomath}
Differentiating the above equation with respect to $\tensor{\gamma}$ gives 
\begin{linenomath}
\begin{align}
  \dfrac{\pd \zeta_{f}}{\pd \tensor{\gamma}} = \dfrac{G \left(\dfrac{\tau_{f}}{\lvert \tensor{\tau}_{f} \cdot \tensor{n} \rvert } \tensor{n} \right)}{\left[ G \alpha \dfrac{\tau_{f}}{\lvert \tensor{\tau}_{f} \cdot \tensor{n} \rvert } \Gamma_d(d, \grad d) + \left(p_{\cn}\dfrac{\dd \mu }{\dd V} + \eta \right) \dfrac{1}{\Delta t} \right]} . 
\end{align}
\end{linenomath}
Finally, we insert the above equation into Eq.~\eqref{eq:tangent-operator-unfinished} and obtain 
\begin{linenomath}
\begin{align}
  \mathbb{C}_{f} = G \left \{\tensor{1} - G \alpha \Gamma_d(d, \grad d) \dfrac{\tau_{f}}{\lvert \tensor{\tau}_{f} \cdot \tensor{n} \rvert }\dfrac{(\tensor{n} \dyadic \tensor{n}) }{\left[ G \alpha \dfrac{\tau_{f}}{\lvert \tensor{\tau}_{f} \cdot \tensor{n} \rvert } \Gamma_d(d, \grad d) + \left(p_{\cn}\dfrac{\dd \mu }{\dd V} + \eta \right) \dfrac{1}{\Delta t} \right]} \right \}.
\end{align}
\end{linenomath}

\section*{Data Availability Statement} 
\label{sec:data-availability} 

The data that support the findings of this study are available from the corresponding author upon reasonable request.

\bibliography{references}

\end{document}